\pdfoutput=1
\documentclass[a4paper,fleqn,usenatbib]{mnras}

\usepackage{amsmath}	
\usepackage{amssymb}	
\usepackage{txfonts}

\usepackage[T1]{fontenc}
\usepackage{ae,aecompl}

\usepackage{graphicx}	
\usepackage{epstopdf}
\usepackage{subcaption}
\usepackage{multicol}
\usepackage{multirow}
\usepackage{float}
\usepackage{enumitem}

\newcommand{\ha}{H$\alpha$}
\newcommand{\hb}{H$\beta+$[O{\sc iii}]}
\newcommand{\oiii}{[O{\sc iii}]}
\newcommand{\oii}{[O{\sc ii}]}
\newcommand{\ioni}{[O{\sc iii}]/[O{\sc ii}]}
\newcommand{\mst}{$M_\mathrm{stellar}$}
\newcommand{\ewr}{EW$_{\mathrm{rest}}$}
\newcommand{\msol}{M$_\odot$}

\title[SMFs and EWs out to $z \sim 5$]{The Nature of \hb~and \oii~emitters to $z \sim 5$ with HiZELS: stellar mass functions and the evolution of EWs}
\author[Khostovan et al.]{A.~A.~Khostovan$^{1}$\thanks{E-mail:
akhostov@gmail.com}, D.~Sobral$^{2,3}$, B.~Mobasher$^{1}$, I.~Smail$^{4,5}$, B.~Darvish$^6$, \newauthor H.~Nayyeri$^7$, S.~Hemmati$^8$, J.~P. Stott$^{9}$\\
$^{1}$Department of Physics \& Astronomy, University of California, Riverside, CA 92512, USA\\
$^{2}$Department of Physics, Lancaster University, Lancaster, LA1 4YB, UK \\
$^{3}$Leiden Observatory, Leiden University, PO Box 9513, NL-2300 RA Leiden, the Netherlands\\
$^{4}$Centre for Extragalactic Astrophysics, Department of Physics, Durham University, Durham DH1 3LE, UK\\
$^{5}$Institute for Computational Cosmology, Durham University, Durham DH1 3LE, UK\\
$^{6}$Cahill Center for Astronomy and Astrophysics, California Institute of Technology, Pasadena, CA 91125, USA\\
$^{7}$Department of Physics \& Astronomy, University of California, Irvine, CA 92697, USA\\
$^{8}$Infrared Processing and Analysis Center, California Institute of Technology, Pasadena, CA 91125, USA\\
$^{9}$Sub-department of Astrophysics, Department of Physics, University of Oxford, Oxford OX1 3RH, UK}
\date{}

\pubyear{2016}

\begin{document}

\label{firstpage}
\pagerange{\pageref{firstpage}--\pageref{lastpage}}
\maketitle

\begin{abstract}
We investigate the properties of $\sim 7000$ narrow-band selected galaxies with strong \hb~and \oii~nebular emission lines from the High-$z$ Emission Line Survey (HiZELS) between $z \sim 0.8 - 5.0$. Our sample covers a wide range in stellar mass ($M_\mathrm{stellar} \sim 10^{7.5 - 12.0}$ \msol), rest-frame equivalent widths (\ewr $\sim 10 -  10^5$ \AA), and line luminosities ($L_\mathrm{line} \sim 10^{40.5 - 43.2}$ erg s$^{-1}$). We measure the \hb-selected stellar mass functions out to $z \sim 3.5$ and find that both $M_\star$ and $\phi_\star$ increases with cosmic time, which may be due to the \oiii~selection including an increasing fraction of AGN at lower redshifts. The \oii-selected stellar mass functions show a constant $M_\star \approx 10^{11.6}$ \msol~and a strong, increasing evolution with cosmic time in $\phi_\star$ in line with \ha~studies. We also investigate the evolution of the \ewr~as a function of redshift with a fixed mass range (10$^{9.5 - 10.0}$ \msol) and find an increasing trend best represented by $(1+z)^{3.81\pm0.14}$ and $(1+z)^{2.72\pm0.19}$ up to $z\sim 2$ and $z \sim 3$ for \hb~and \oii~emitters, respectively. This is the first time that the \ewr~evolution has been directly measured for \hb~and \oii~emitters up to these redshifts. There is evidence for a slower evolution for $z > 2$ in the \hb~\ewr~and a decreasing trend for $z > 3$ in the \oii~\ewr~evolution, which would imply low \oii~EW at the highest redshifts and higher \ioni~line ratios. This suggests that the ionization parameter at higher redshift may be significantly higher than the local Universe. Our results set the stage for future near-IR space-based spectroscopic surveys to test our extrapolated predictions and also produce $z > 5$ measurements to constrain the high-$z$ end of the \ewr~and \ioni~evolution.

\end{abstract}


\section{Introduction}
In the past two decades, great strides have been made in understanding the evolution of observed properties of star-forming galaxies across cosmic time. We now know that the peak of star formation activity occurred somewhere between $z \sim 2$ and $3$ (e.g., \citealt{Karim2011,Bouwens2012a,Bouwens2012b,Cucciati2012,Gruppioni2013,Sobral2013,Bouwens2015,Khostovan2015}) and that the majority of the stellar mass assembly occurred by $z \sim 1$ (e.g., \citealt{Perez-Gonzalez2008,Marchesini2009,Ilbert2013,Muzzin2013,Madau2014,Sobral2014,Tomczak2014,Grazian2015}). Furthermore, recent spectroscopic surveys are giving us valuable insight on the physical properties of star-forming regions in the high-$z$ Universe (e.g., \citealt{Liu2008,Swinbank2012,Nakajima2013,Sobral2013b,Nakajima2014,Newman2014,Shirazi2014,Steidel2014,Stott2014,Hayashi2015,Sanders2016,Stott2016}). 

As galaxies age and undergo star-formation, the byproduct of their star-formation activity is their stellar mass build up. Therefore, determining and understanding the evolution of the stellar mass function (SMF) is crucial as measuring the distribution of stellar mass within a given comoving volume provides important observational evidence on how galaxies may grow due to star formation (e.g., \citealt{vanDokkum2010, Bauer2013}), mergers (e.g., \citealt{Drory2008, Vulcani2015}), and environmental influences (e.g., \citealt{Baldry2006, Bundy2006, Bolzonella2010, Peng2010, Sobral2011, Giodini2012, Darvish2015a, Mortlock2015, Davidzon2015, Sobral2016}). Measurements of the SMFs also provide valuable constraints for theoretical models of the hierarchical assembly of dark matter halos (e.g., SMF-DM Halo Mass relationship; \citealt{Conroy2009, Behroozi2013,Furlong2015,Henriques2015,Rodriguez2015}; for a recent review article see \citealt{Somerville2015}). 

Another observational tracer of galaxy formation and evolution is the stellar mass density (SMD), which measures the total stellar mass within a specific range of masses (e.g., $> 10^9$ \msol) or full range (e.g., integrating the SMF from zero to infinity) per unit of comoving volume. By combining with other SMD measurements over a wide redshift range, the evolution of the SMD can be measured and reveal how galaxies assembled their stellar mass over cosmic time. In a compilation of the latest SMD measurements (e.g., \citealt{Arnouts2007, Gallazzi2008, Perez-Gonzalez2008, Kajisawa2009, Li2009, Marchesini2009, Yabe2009, Pozzetti2010, Caputi2011, Gonzalez2011, Bielby2012, Lee2012, Reddy2012, Ilbert2013, Moustakas2013, Muzzin2013, Labbe2013}), \citet{Madau2014} showed a strong, increasing trend from $z \sim 8$ to $z \sim 1$, followed by a shallower, increasing trend from $z \sim 1$ to the present. This evolution is correlated with the cosmic star-formation rate density (SFRD) evolution, such that it is possible to model the SMD evolution based on the average SFRD evolution by taking its time integral (e.g., \citealt{Sobral2013,Madau2014, Khostovan2015}) and vice versa via the time derivative (e.g., \citealt{Perez-Gonzalez2008}). 

Despite the various measurements that have provided a general indication of the SMF and SMD evolution, there are several caveats. For example, spectral energy distribution (SED) models and templates used to measure stellar masses can introduce systematic biases based on assumptions made in the fitting process and differing methodologies \citep{Mobasher2015}. Also, the separation based on galaxy types typically is based on empirically-derived colour-colour selection diagnostics (e.g., $BzK$, \citealt{Daddi2004}; $UVJ$, \citealt{Williams2009}), which can vary based on the data-set used (e.g., selection effects arising from sample and/or survey size and depth). Therefore, to make further progress we need a reliable, clean sample of a specific type of galaxies over a large comoving volume that can trace the SMF and SMD evolution from low-$z$ to high-$z$ using a single methodology.

Recently, there has been a great deal of focus on the evolution of the specific star-formation rate (sSFR), which is defined as the star-formation rate divided by the stellar mass (e.g., \citealt{Stark2013, Gonzalez2014,Marmol2015,Faisst2016}). Since the sSFR is in inverse units of time, it can be interpreted as a direct measurement of the timescale of stellar growth in individual galaxies and also as the ratio between the current and past star-formation activity. Recent studies have constrained the evolution within the $z < 2$ regime, finding that the sSFR increases from $z = 0$ to $z \sim 2$ \citep{Noeske2007,Damen2009,Sobral2014}. 

However, the sSFR evolution is less constrained for $z > 2$. \citet{Reddy2012} measured the sSFR evolution between $z \sim 2 - 3$ and \citet{Stark2009} and \citet{Gonzalez2010} extended the measurements for $z > 4$. In comparison with the $z < 2$ data, the observational data show the sSFR increasing from $0.3$ to $2$ Gyr$^{-1}$ between $z = 0 $ and $z \sim 2$. For $z > 2$, some early studies found that sSFR showed no significant evolution and is claimed to stay flat around $\sim 2$ Gyr$^{-1}$ up to $z \sim 7$ \citep{Stark2009,Gonzalez2010}. In contrast, theoretical studies predict that, for the case of cold gas accretion growth, the sSFR should be increasing as $(1 + z)^{2.25}$ \citep{Dave2011, Dave2012}. Latest measurements from the high-resolution EAGLE simulation also predict an increasing sSFR with redshift \citep{Furlong2015}. An issue that can arise for the observational studies at $z > 4$ is that they do not take into account the effects of nebular emission lines in the SED fitting process. Strong lines can contaminate the {\it Spitzer} IRAC bands at these redshifts resulting in overestimating stellar masses (e.g., \citealt{Schaerer2009,Schaerer2010,Nayyeri2014, Smit2014}). Looking at $\sim 1700$ $z \sim 3 - 6$ Lyman break galaxies (LBGs), \citet{deBarros2014} found that about two thirds of their sample had detectable emission lines and by taking them into account when fitting the SED resulted in significantly different physical parameters. Recently, \citet{Gonzalez2014} presented newer measurements of the sSFR with the nebular contamination accounted for and found an increase of a factor of $\sim 2$ in comparison to the \citet{Stark2009} and \citet{Gonzalez2010} measurements, but still in conflict with theoretical predictions.

To correct the overestimation of stellar masses and sSFRs requires that the contamination of nebular emission lines is taken into account. One way of doing this is by measuring the trends in the rest-frame equivalent widths (\ewr) of lines, which is a ratio between the flux of the emission-line and the stellar continuum flux. Studies have mapped out the \ewr(H$\alpha$) evolution up to $z \sim 2$ (e.g., \citealt{Erb2006, Fumagalli2012, Sobral2014}) while the $z > 2$ trend is still uncertain since H$\alpha$ falls into the infrared at these redshifts. Recent measurements, using colour excess in the {\it Spitzer} IRAC bands at $> 3$\micron~that are claimed to only be attributed to nebular emission line contribution, have attempted to extend the measurements of the evolution out to $z \sim 6$ (e.g., \citealt{Shim2011, Rasappu2015}). Other studies measured \ewr(\hb) between $z \sim 6 - 8$ and, using a line ratio, converted to H$\alpha$ to extend the mapping of the \ewr(H$\alpha$) evolution (e.g., \citealt{Labbe2013, Smit2014, Smit2015}). It should be noted that current studies are UV selected and are only sensitive to the most extreme line-emitters which can be detected in the broad-band photometry. Therefore, these measurements can be only treated as upper-limits. What we require are complete samples of emission-line selected sources (e.g., cover a wide-range in \ewr~that represents a typical emission-line galaxy) to properly measure the \ewr~evolution at $z > 2$. The lines that can be used are \hb\footnotemark~up to $z \sim 3$ and \oii~up to $z \sim 5$ (e.g., \citealt{Khostovan2015}).

\footnotetext{The narrow-band filters used all have FWHMs of $\sim 130 - 210$\AA~and can differentiate between H$\beta$ and \oiii~emitters, but the broad-band filters used in selecting sources have FWHMs too large to separate the sample. Therefore, our \hb~samples are comprised of a combination of H$\beta$ and \oiii, although \citet{Khostovan2015} and \citet{Sobral2015} showed that the samples are dominated by \oiii~emitters.}

Tracing the evolution of the equivalent width of nebular emission lines also provides valuable insight to the physical conditions of the H{\sc ii} regions and how those physical conditions evolve over cosmic time (e.g., \citealt{Liu2008, Nakajima2013, Nakajima2014, Hayashi2015, Kewley2015}). For example, the \ioni~ratio as measured by \ewr(\oiii)/\ewr(\oii) can, in principle, tell us about the ionization parameter and the ionization state of the gas forming stars.

In this paper, we present our investigation of the evolution in SMF, SMD, and \ewr~using a large sample of \hb~and \oii~emission-line galaxies at $z \sim 1 - 5$ from the High-Redshift Emission-Line Survey (HiZELS) presented by \citet{Khostovan2015}. Our results have implications in terms of the evolution in the \ewr~and sSFR, as well as the physical conditions of the gas in the H{\sc ii} regions that produces the nebular emission-lines. Our results also present an empirical evolution of the \ewr~that can be used to estimate the nebular emission line contamination in broad-band photometry when such photometry are used in determining key physical properties (e.g., stellar masses). 

The paper is structured as follows: Section \ref{sec:sample} describes the HiZELS sample used in this paper; Section \ref{sec:methodology} presents the stellar mass, SMF, and SMD determinations; Section \ref{sec:results} highlights the results of this paper with interpretations of the SMF, SMD, \ewr, and \ioni~evolutions; Section \ref{sec:conclusion} summarizes the main results of our study.

Throughout this paper, we assume $\Lambda$CDM cosmology, with $H_0 = 70$ km s$^{-1}$ Mpc$^{-1}$, $\Omega_\Lambda = 0.3$, and $\Omega_\mathrm{m} = 0.7$. We assume a \citet{Chabrier2003} IMF and correct the literature measurements when needed. All magnitudes are presented as AB magnitudes \citep{Oke1983}.

\section{HiZELS Sample}
\label{sec:sample}

Our sample consists of \hb~and \oii~emitters selected based on narrow-band photometry from HiZELS \citep{Geach2008, Sobral2009, Sobral2012, Best2013, Sobral2013} found in the COSMOS \citep{Scoville2007} and UDS \citep{Lawrence2007} fields. We refer the reader to \citet{Sobral2013} for details on the initial selection of sources with narrow-band excess. 

The sample consists of 3475 \hb~emitters between $z = 0.84$ and 3.24 and 3298 \oii~emitters between $z = 1.47$ and 4.69 in discrete redshift slices (see Table \ref{table:cut}) with the redshifts corresponding to the narrow-band filters used by \citet{Sobral2013}\footnotemark. Our sample is backed by 233 and 219 spectroscopic measurements for \hb~and \oii, respectively, that are from zCOSMOS \citep{Lilly2007}, the UDSz Survey (\citealt{Bradshaw2013, McLure2013}), Subaru-FMOS \citep{Stott2013}, Keck-DEIMOS/MOSFIRE (Nayyeri et al., in prep), PRIsm MUlti-object Survey (PRIMUS; \citealt{Coil2011}), and VIMOS Public Extragalactic Redshift Survey (VIPERS; \citealt{Garilli2014}). This sample is based on a large areal coverage of $\sim 2$ deg$^{2}$ equating to a comoving volume coverage of $\sim 10^6$ Mpc$^{3}$, which greatly reduces the effects of cosmic variance (see \citealt{Sobral2015,Stroe2015}). 

\footnotetext{We refer the reader to \citet{Sobral2013} for information regarding the filter profiles, FWHMs, effective wavelengths, and all other inquiries regarding the properties of the narrow-band and broad-band filters used.}

The selection of \hb~and \oii~emitters is discussed in \citet{Khostovan2015}. In brief, we used the emission-line source catalog of \citet{Sobral2013} to select galaxies with \hb~or \oii~emission lines by using a combination of selection criteria: spectroscopic redshifts, photometric redshifts, and colour-colour diagnostics (with priority given in that order). Sources that had detections in more than one narrow-band filter were also selected on the basis that their confirmation is equivalent to spectroscopic confirmation (e.g., finding \oii~in NB921 and \ha~in NBH at $z = 1.47$; see \citealt{Sobral2012}).

\begin{figure*}
\centering
\includegraphics[width=0.95\textwidth]{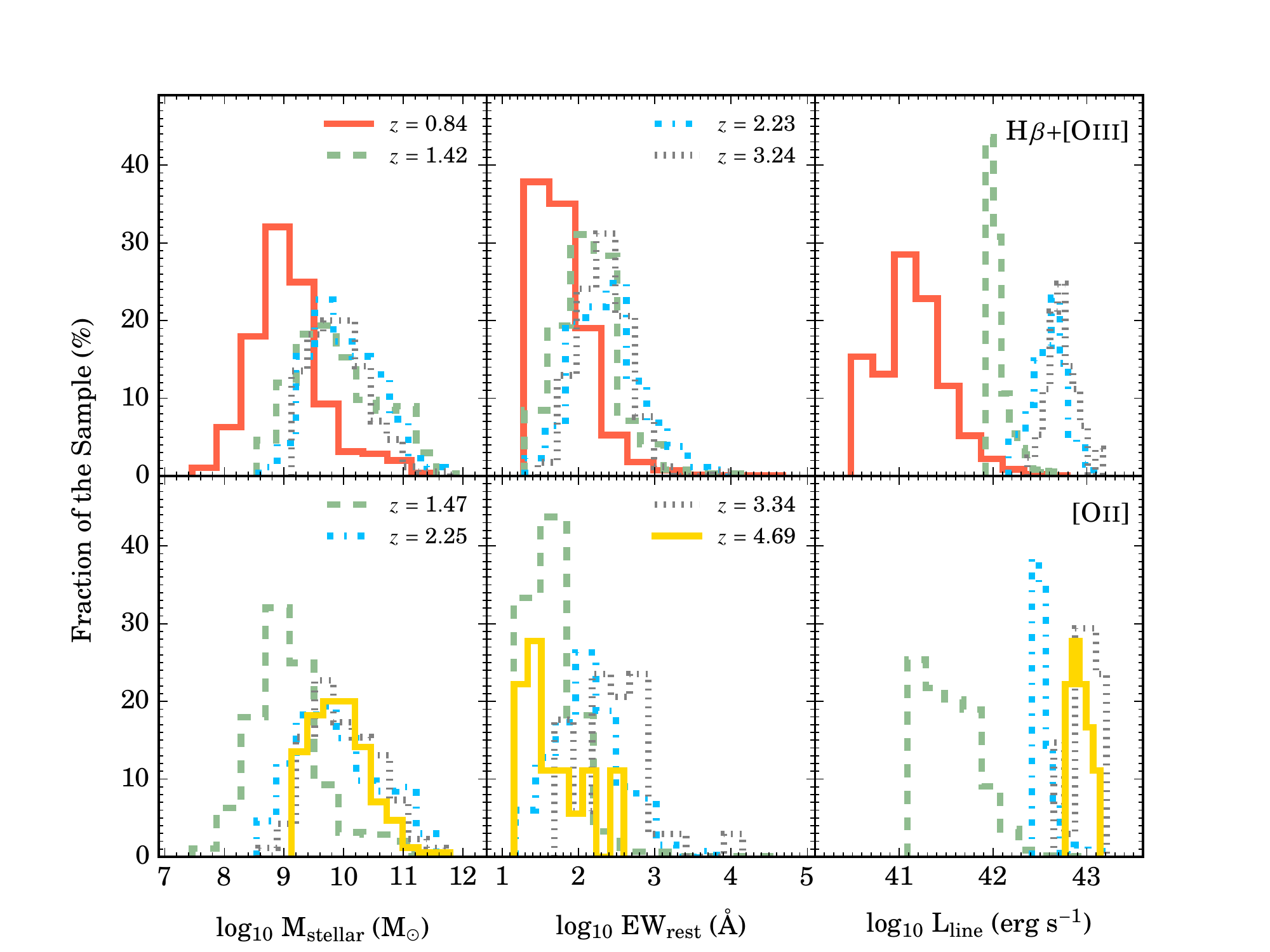}
\caption{The stellar mass, \ewr, and luminosity distributions for all of our samples. Based on the luminosity distributions, it is clear that our high-$z$ sample is limited to high line luminosities ($L > 10^{42}$ erg s$^{-1}$). Our lowest redshift sample is the deepest and covers a wider luminosity, stellar mass, and \ewr~range which allows us to utilize the sample for tests of selection effects that can bias results at higher redshift.}
\label{fig:hist}
\end{figure*}

The rest-frame equivalent widths of emission lines are calculated using the following relation:
\begin{equation}
\mathrm{EW}_\mathrm{rest} \approx \frac{F_L}{f_C} = \frac{\Delta \lambda_{\mathrm{NB}}}{1 + z} \frac{f_\mathrm{NB} - f_\mathrm{BB}}{f_\mathrm{BB} - f_\mathrm{NB}(\Delta \lambda_\mathrm{NB}/\Delta \lambda_\mathrm{BB})}
\label{eqn:ewr}
\end{equation}
where NB and BB are the narrow-band and broad-band filters, respectively, $\Delta \lambda$ is the corresponding width of the filter, $f$ is the corresponding flux measured in the filter, $F_L$ is the flux of the nebular emission line, and $f_C$ is the continuum flux. Figure \ref{fig:hist} shows the distribution of rest-frame EWs and line luminosities of the \hb~and \oii~emitters. Our sample consists of rest-frame equivalent widths that are as low as $\sim 10$ \AA~and as high as 10$^5$ \AA~and a luminosity range between 10$^{40.5}$ and 10$^{43.2}$ erg s$^{-1}$. We refer the reader to \citet{Khostovan2015} for details on how the line luminosities were computed.

\section{Methodology}
\label{sec:methodology}

\subsection{SED Fitting}

We use the Multi-wavelength Analysis of Galaxy PHYSical properties ({\sc MAGPHYS}) code of \citet{Cunha2008} to fit the SEDs of our sources and determine physical properties, such as stellar masses, star-formation rates, and $E(B-V)$. \citet{Cunha2008} designed the code to treat the infrared as two sub-components (birth clouds and diffuse ISM) using empirical relations from \citet{Charlot2000} and assuming a balance between the stellar and dust/infrared components (e.g., the amount of attenuation in the stellar component is accounted for in the dust/infrared component). 

{\sc MAGPHYS} uses different model templates for the stellar and infrared components. The stellar component is generated by the \citet{Bruzual2003} code, while the infrared component templates are formed based on the prescription of \citet{Charlot2000}. We note that {\sc MAGPHYS} assumes a \citet{Chabrier2003} IMF\footnotemark. The stellar templates include (1) exponentially declining star-formation histories $e^{-t/\tau}$ with $\tau$ in the range between 0.1 to 13.5 Gyr; (2) metallicities between 0.02 and 2 $Z_\odot$; and (3) dust attenuation based on \citet{Charlot2000}. {\sc MAGPHYS} then fits the observed SEDs and creates marginalized likelihood distributions of physical parameters.

\footnotetext{To make our results comparable with other studies in the literature that utilize different IMFs, we state the conversions to the Salpeter IMF ($+0.215$ dex) and the Kroupa IMF ($-0.04$ dex).}

We fit the SEDs using {\it GALEX} $FUV$ and $NUV$, CFHT Megaprime $u^*$, Subaru SuprimeCam $Bg'Vr'i'z'$, UKIRT WFCAM $J$ and $K$, and {\it Spitzer} IRAC 3.6 - 8.0 \micron~photometry for our COSMOS sources. The SEDs of our UDS sources are fitted using CFHT MegaCam $u$, Subaru SuprimeCam $BVr'i'z'$, UKIRT WFCAM $YJHK$, and {\it Spitzer} IRAC 3.6 - 8.0 \micron~photometry. The outputs used in this study are the stellar masses. We use the COSMOS-30 $i$-band selected catalog \citep{Capak2007, Ilbert2009} and the DR8 release of the Subaru-XMM-UKIDSS UDS $K$-band selected catalog (e.g., \citealt{Cirasuolo2007, Lawrence2007}). We refer the reader to the cited catalog papers for detailed descriptions of the multi-wavelength photometry.

We note that {\sc MAGPHYS} was created to incorporate the $912$ \AA~$< \lambda < 1$ mm rest-frame range such that we have no mid- and far-infrared constraints. The unique part about {\sc MAGPHYS} is that it fits the stellar and infrared/dust templates separately, such that in the case where there are no infrared constraints, the measurements will be based off of the fits using only the stellar templates. Furthermore, {\sc MAGPHYS} does not have a prescription to incorporate the effects of nebular emission in the fitting process. Past studies have shown that nebular emission contamination can affect the stellar mass measurements from SED fitting (e.g., \citealt{Schaerer2009,deBarros2014}). As shown in Figure \ref{fig:hist}, we find that for the vast majority of our sources, their \ewr~are low enough ($< 10^3$ \AA; e.g., \citealt{Smit2014}) and have $\sim 10 - 15$ individual photometric measurements for which the effects of nebular emission line contamination are negligible.

\subsection{Stellar Masses}

Figure \ref{fig:hist} shows the range in stellar mass that were measured from {\sc MAGPHYS} for all our samples. We find that our $z > 1$ \hb~emitters and $z > 1.5$ \oii~emitters have typical stellar masses \mst$\sim 10^{9.5} - 10^{10}$ \msol. The $z = 0.84$ \hb~and $z = 1.47$ \oii~samples have distributions that peak at lower masses (\mst$\sim 10^{8.5} - 10^9$ \msol) and cover a wider range (\mst$\sim 10^{7.5} - 10^{12.0}$\msol). Both samples come from NB921 observations, which, as seen in the luminosity distributions shown on Figure \ref{fig:hist}, probe deeper than all the other samples, but also covers a much smaller volume ($\sim 3 - 7 \times 10^5$ Mpc$^{-3}$, \citealt{Khostovan2015}). Since the COSMOS field has a wealth of multi-wavelength with measurements of stellar masses, we make a comparison between our measurements and those of \citet{Ilbert2010} and \citet{Muzzin2013} as shown in Appendix \ref{sec:comparison}. We find that our measurements are consistent with those of the literature.

\subsection{Creating Stellar Mass Functions}
\label{sec:make_SMF}

We create stellar mass functions by using a similar approach as in \citet{Khostovan2015} by applying the V$_\mathrm{max}$ estimator where the data is binned as such:
\begin{equation}
\phi(M_j) = \frac{1}{\Delta M_j} \sum_{i=0}^{N} \frac{1}{C(M_i) V_{\mathrm{max},i}}
\end{equation}
where $M_j$ is the $j^{\mathrm{th}}$ mass bin, $\Delta M_{j}$ is the bin-size, and $C(M_i)$ is the completeness and $V_{\mathrm{max,i}}$ is the volume for the $i^{\mathrm{th}}$ source in the $j^{\mathrm{th}}$ bin. The masses, $M$, used in this equation are all in log-scale. 

\subsubsection{Completeness Correction}
All the stellar mass functions have been corrected for completeness based on the completeness corrections determined by \citet{Khostovan2015} using the approach of \citet{Sobral2013, Sobral2014}. We adopt this approach for correcting our SMFs because our samples are flux- and EW-limited and not mass-limited. Therefore, we need to correct based on the line flux and EWs as this is where the incompleteness arises. In brief, the completeness correction takes into account the full selection function (including the \ewr~cut and the difference in luminosity limits/depths between one sub-field and another) in terms of line luminosity. Furthermore, we applied a volume/filter profile correction (see \citealt{Khostovan2015}), which takes into account the loss of flux at the wings of the narrow-band filters. We also applied an \ewr~completeness correction to take into account the missing number of high mass galaxies in our $z = 0.84$ SMF. This is described in detail in Appendix \ref{sec:ewr_comp}.

\subsubsection{Common Relative Luminosity Cut}
As seen in Figure \ref{fig:hist}, each sample covers different line luminosities making it difficult to directly compare samples. Furthermore, the volumes probed per each sample are different where the lowest redshift samples have comoving volumes of $\sim 3 - 6 \times 10^5$ Mpc$^{3}$ and the highest redshift samples with $\sim 10 - 16 \times 10^{5}$ Mpc$^{3}$ \citep{Khostovan2015}. This raises problems in terms of compatibility for comparison as the line luminosity and volume differences can capture different populations of galaxies. In order to solve this issue, we use a similar approach to \citet{Sobral2014} by placing a common $L/L_\star(z)$ limit to make the samples directly comparable using the $L_\star(z)$ measurements of \citet{Khostovan2015}. This is accomplished by comparing the distribution of sources per redshift in terms of their $L/L_\star(z)$ ratio where we find that the common limit for \hb~is $\sim 0.4 L_\star(z)$ and for \oii~it is $\sim 0.85 L_\star(z)$ (see Table 3 of \citealt{Khostovan2015} for the $L_\star(z)$ measurements). Disregarding this common limit will result in stellar mass functions and densities that trace different types of emitters. 

This consequentially reduces the sample size, especially for the lowest $z$ samples ($z \sim 0.84$ for \hb~and $z \sim 1.47$ for \oii) as they are the deepest and have the largest sample size. Table \ref{table:cut} shows the change in sample size when applying the common relative luminosity cut. Percentages shown correspond to the percentage of sources that were selected in comparison to the full sample. The NB921 samples (\hb~$z = 0.84$ and \oii~$z = 1.47$) saw the largest reductions in sample size due to their line luminosity distributions peaking at lower luminosities (see Figure \ref{fig:hist} for the line luminosity distributions of all the samples). The higher-$z$ samples retain the vast majority of their original sample sizes due to the fact that the lower $L/L_\star(z)$ limit chosen was on their line luminosity distributions.

\begin{table}
\centering
\begin{tabular}{ccccc}
\hline
$z$ & $L_\star$ (erg s$^{-1}$) & $N_\mathrm{total}$ & $N_\mathrm{sel}$ & Fraction \\
 \hline
 \multicolumn{5}{c}{\hb~($L > 0.4 L_\star(z)$)}\\
 \hline
 0.84 & 41.79 & 2477 & 524 & 21\% \\
 1.42 & 42.06 & 371 & 371 & 100\% \\
 2.23 & 42.66 & 271 & 256 & 95\% \\
 3.24 & 42.83 & 179 & 175 & 98\% \\
 \hline
 \multicolumn{5}{c}{\oii~($L > 0.85 L_\star(z)$)}\\
 \hline
 1.47 & 41.86 & 3285 & 676 & 21\% \\
 2.25 & 42.34 & 137 & 137 & 100\% \\
 3.34 & 42.69 & 35 & 35 & 100\% \\
 \hline
 \end{tabular}
 \caption{To ensure compatibility between different redshift samples, we apply a common $L/L_\star(z)$ cut (\hb: $L > 0.4 L_\star(z)$ and \oii: $L > 0.85 L_\star(z)$ where the $L_\star(z)$ measurements are from \citet{Khostovan2015}. For each redshift sample, we highlight the total number of emitters in the sample ($N_\mathrm{total}$), the total number of emitters selected after the $L/L_\star(z)$ cut ($N_\mathrm{sel}$), and the corresponding fraction of emitters selected.}
 \label{table:cut}
 \end{table}

\subsection{Which one dominates: H$\beta$ or \oiii?}
\label{sec:hb_o3}
As mentioned in \S \ref{sec:sample} and discussed in \citet{Khostovan2015}, our \hb~sample is a combination of H$\beta$ and \oiii~emitters. The narrow-band filters can differentiate between the different emission lines. The problem arises in the selection techniques used by \citet{Khostovan2015}, which, as briefly described in \S \ref{sec:sample}, rely on a combination of spectroscopic confirmation, photometric redshifts, and colour-colour criteria. The photometric redshifts and colour-colour criteria both depend on using the multi-wavelength broad-band filters data sets, which results in the H$\beta$ and \oiii~emitters to be blended with each other.

The important question that arises from this is which one dominates the \hb~sample: H$\beta$ or \oiii~emitters? \citet{Khostovan2015} showed that the \oiii~line dominates the population of emitters with the fraction of H$\beta$ emitters increasing with decreasing \hb~line luminosities. In a similar study, \citet{Sobral2015} used their CF-HiZELS $z\sim 1.4$ sample from the $\approx 10$ deg$^2$ SA22 field and found that $\sim 16\%$ of their spectroscopically confirmed \hb~emitters were H$\beta$ emitters. Therefore, we can safely assume that the \oiii~emitters dominate our \hb~sample.

\subsection{Contamination from AGNs}
AGNs will also be selected with narrow-band surveys as the energetic UV photons they release can produce the emission lines that are also produced by the UV photons from bright, massive stars in star-forming, H{\sc ii} regions. \citet{Khostovan2015} and \citet{Sobral2015} both studied the AGN contamination in their samples by using the $1.6\micron$ bump as an observational proxy. Both found, on average, that the AGN contamination is $\sim 10 - 20\%$ of the total population. \citet{Khostovan2015} also compared the \hb~luminosity functions to the $z \sim 0.7$ {\it z}COSMOS \oiii~type-2 AGN luminosity function of \citet{Bongiorno2010} and found that the brightest emitters in the \hb~sample are probable AGNs and as the \hb~line luminosity decreases, so does the fraction of AGN contribution.

We note that any type 1 (broad line) AGN in our sample may result in a poor $\chi^2$ SED fits making them easier to remove from the sample. The type 2 (narrow line) AGNs are harder to remove but can still result in poor $\chi^2$ fits. To remove this contamination, we incorporate a $\chi_\mathrm{reduced}^2 < 100$. 

\section{Results}
\label{sec:results}

\subsection{Quiescent Population?}
\label{sec:UVJ}

\begin{figure}
\centering
\includegraphics[width=1.1\columnwidth]{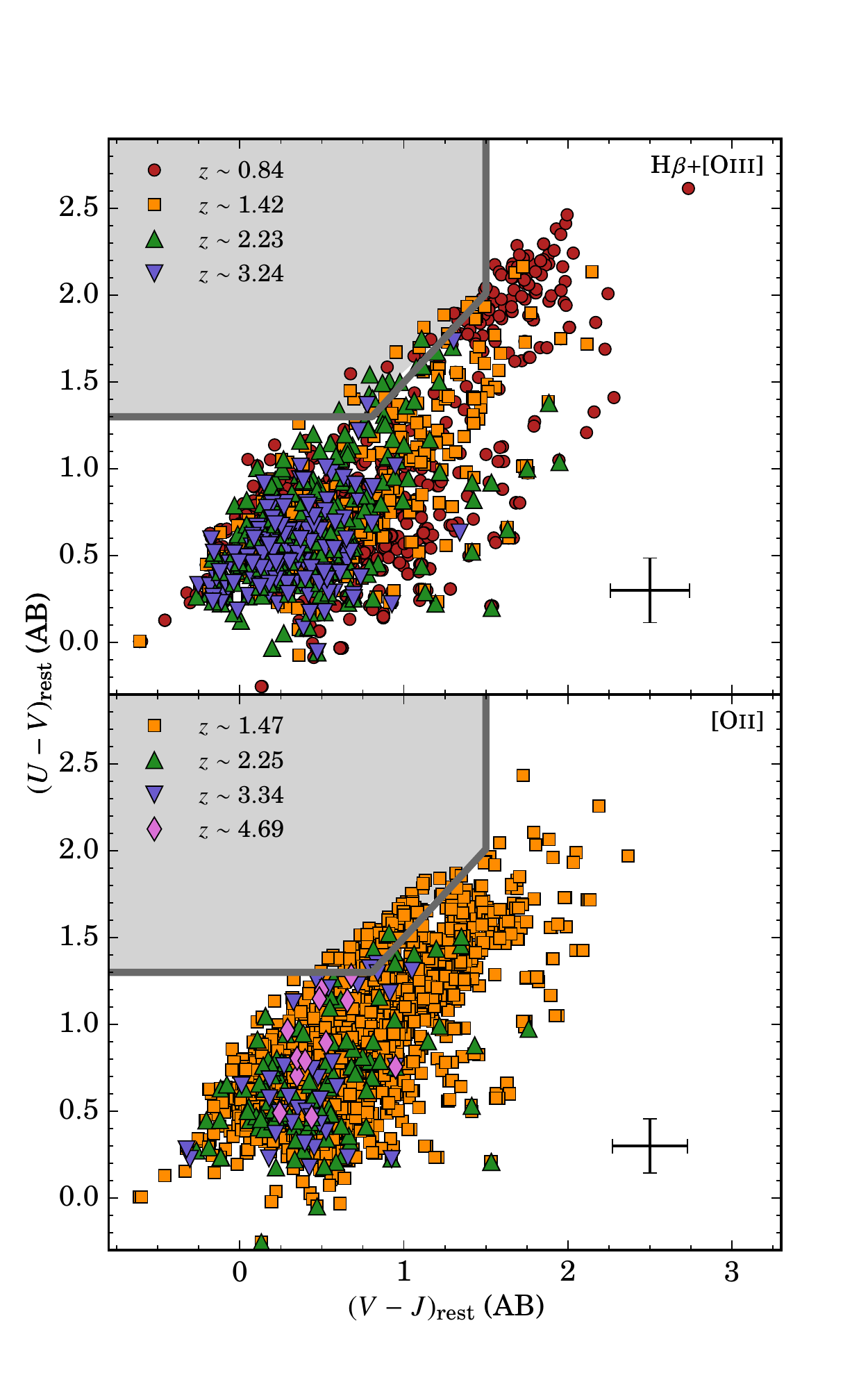}
\caption{The $UVJ$ colour-colour diagnostic used to separate star-forming galaxies from quiescent galaxies. The highlighted region and {\it grey} boundaries are the \citet{Muzzin2013} quiescent selection region. Included is the typical 1$\sigma$ range for all sources per emission line. We find that the majority of our sources reside within the star-forming classification region. Sources that are within the quiescent region are consistent with photometric scatter. We find a general trend in the $UVJ$ plane where high-$z$ sources tend to have bluer rest-frame $UVJ$ colours that could be caused by changes in dust and/or the star-formation efficiency (e.g., \citealt{Papovich2015}).}
\label{fig:UVJ}
\end{figure}

In the past, many studies used the rest-frame $UVJ$ colour-colour selection to separate quiescent/passive and star-forming galaxies (eg., \citealt{Williams2009,Brammer2011,Muzzin2013}). Unobscured star-forming galaxies will have bluer rest-frame $U-V$ colours, corresponding to younger stellar populations and a lower or no 4000\AA~break, and also have bluer $V-J$ colours forming a locus within the $UVJ$ plane. Dust-free quiescent galaxies are dominated by a more evolved stellar population resulting in a more pronounced 4000\AA~break, resulting in redder $U - V$ colours, although dust-obscured star-forming galaxies can occupy the same regime due to attenuation. This degeneracy is broken by $V - J$, where dust-free quiescent galaxies have bluer colours than the dust-obscured star-forming galaxies. The $UVJ$ classification scheme does not take into account possibility of AGN contamination, such that galaxies that fall under both classifications can also be potential AGNs. Both classifications can also include sources with more complex spikes of obscured/unobscured star formation. We therefore refer to the quiescent classification as ``passive" and the star-forming classification as ``active" to take into account AGNs.

It must be noted that the $UVJ$ selection is empirically driven and varies based on the data-set used, as well as the filters used in determining the rest-frame AB magnitudes. We apply the \citet{Muzzin2013} $UVJ$ selection and use the same filters (Johnson $U$ and $V$ and 2MASS $J$) to study the nature of our sample. Figure \ref{fig:UVJ} shows our full sample of emitters and the \citet{Muzzin2013} colour-colour selection. We include the 1$\sigma$ range for all sources per emission line that is calculated from the observed error bars of the corresponding $UVJ$ observer-frame filters.

We find that, for all our \hb~and \oii~samples, $> 98.5\%$ are classified as active based on this selection criteria. We also find a small population of emitters that fall under the passive classification area. For the \hb~sample, only $0.8\%$ (26 emitters) fall within this selection area with the majority ($\sim 38\%$ of the 26 emitters) being from the $z \sim 1.42$ sample. The \oii~sample has a total of $2.4\%$ (79 emitters) of the full sample within the passive selection region with the vast majority ($\sim 96\%$; 76 of the 79 emitters) coming from the $z \sim 1.47$ sample. These are mostly faint sources that fall into the passive selection region and are consistent with photometric scatter. Overall, the sources discussed above make a small fraction of our full sample. 

The $UVJ$ selection criteria also confirms that the great majority ($> 98\%$) of our sample can be classified as active. There is also a general trend where rest-frame colours become bluer with increasing redshift implying that our high-$z$ samples are likely comprised of less dusty systems. This could be attributed to sample bias as dusty systems would result in fainter emission-line fluxes leaving behind the less dusty and observationally bright systems (e.g., \citealt{Hayashi2013}). This leads to the caveat that the samples are not fully comparable across redshift as we will be missing the dustier systems. On the other hand, this could also indicate that there is a redshift evolution in the $UVJ$ plane for which galaxies at high-$z$ tend to have bluer rest-frame colours. As these galaxies evolve and their star-formation efficiency decreases and the amount of dust increases, their $UVJ$ colours become redder. This is consistent with the Milky Way progenitor evolution study of \citet{Papovich2015}.

\subsection{Stellar Mass Functions}
In this section, we present the stellar mass function (SMF) of line emitters up to $z \sim 3$ (we exclude the $z = 4.69$ \oii~emitters since we could not constrain the SMF due to the small sample size). All samples used to measure the SMF have a common $L/L_\star(z)$ cut ($0.4 L_\star(z)$ and $0.85 L_\star(z)$ for \hb~and \oii, respectively) in order to make them comparable (tracing a similar galaxy population). The observed measurements are shown in Figures \ref{fig:SMF}. All the measurements have been completeness and filter profile corrected as described in \S \ref{sec:make_SMF}. We fit the observed binned data to the Schechter function in log-form:
\begin{equation}
\Phi(M) \mathrm{d} M = \phi_\star ~\ln 10 ~\Bigg (\frac{M}{M_\star}\Bigg)^{1+\alpha} e^{-(M/M_\star)}\mathrm{d} \log_{10} M
\end{equation}
where $\phi_\star$ is the normalization, $M_\star$ is the characteristic mass, and $\alpha$ is the faint-end slope. The fits are plotted in Figures \ref{fig:SMF} with the fitted parameters shown in Table \ref{table:params}. Note that we also placed a $L/L_\star(z)$ limit as discussed at the end of \S\ref{sec:make_SMF} to make all our samples comparable to one another (tracing the same type of emitters). 

We initially measure the faint-end slope for our deepest samples and compare them to those measured in the literature \citep{Perez-Gonzalez2008,Marchesini2009,Muzzin2013}. Based on these three studies that trace the SMF evolution up to $z \sim 5$, the faint-end slope does not evolve strongly. Therefore, to be comparable from sample to sample and also to the literature when making our comparisons, we fix $\alpha = -1.3$ and refit for $\phi_\star$ and $M_\star$ (shown in Table \ref{table:params}). Note that our measured $\alpha$ for the \hb~$z = 0.84$ and $z = 1.42$ SMFs and the \oii~$z = 1.47$ SMF are in agreement with the fixed $\alpha$ constraint as shown in Table \ref{table:params}.

\begin{table*}
\centering
\caption{Our fitted Schechter parameters of our stellar mass functions. Shown are the parameters when $\alpha$ is free and also in the case when we fix $\alpha$ to $-1.3$ in order to make our measurements comparable with the literature. Note that only the $z = 0.84$ and $1.42$ \hb~measurements and $z = 1.47$ \oii~measurements are used for the case of $\alpha$ being free. This is because the sample size was large enough to probe the faint-end slope, which is then used in comparison to the literature to set a fixed $\alpha$ for all redshift samples. We show the Schechter parameters for where $\alpha$ is free only for our most populated samples. Stellar mass densities are calculated by fully integrating the stellar mass functions. Also included is the $L/L_\star(z)$ limit used to make all the samples compatible for comparison.}%
\begin{tabular}{lcccccc}
\hline
\multicolumn{7}{c}{\hb-selected Stellar Mass Function Properties ($L/L_\star(z) > 0.4$)}\\
\hline
$z$ & $\log_{10} \phi_\star$ & $\log_{10} M_\star$ & $\alpha$ & $\log_{10} \phi_{\alpha = -1.3}$ & $\log_{10} M_{\star,\alpha = -1.3}$ & $\log_{10} \rho_{\star,\alpha = -1.3}$\\
& (Mpc$^{-3}$) & (M$_\odot$) &  & (Mpc$^{-3}$) & (M$_\odot$) & (M$_\odot$ Mpc$^{-3}$)\\
\hline
0.84 & $-3.77_{-0.20}^{+0.16}$ & $11.49_{-0.17}^{+0.30}$ & $-1.27_{-0.07}^{+0.06}$ & $-3.87_{-0.11}^{+0.06}$ & $11.60_{-0.13}^{+0.29}$ & $7.62_{-0.08}^{+0.20}$\\
1.42 & $-3.88_{-0.16}^{+0.12}$ & $11.53_{-0.09}^{+0.17}$ & $-1.28_{-0.08}^{+0.07}$ & $-3.90_{-0.07}^{+0.05}$ & $11.55_{-0.08}^{+0.12}$ & $7.76_{-0.06}^{+0.07}$\\
2.23 & ... & ... & ... & $-4.16_{-0.07}^{+0.05}$ & $11.22_{-0.07}^{+0.11}$ & $7.18_{-0.05}^{+0.06}$\\
3.24 & ... & ... & ... & $-4.16_{-0.08}^{+0.08}$ & $10.96_{-0.08}^{+0.15}$ & $6.90_{-0.06}^{+0.07}$\\
\hline
\multicolumn{7}{c}{\oii-selected Stellar Mass Function Properties ($L/L_\star(z) > 0.85$)}\\
\hline
1.47 & $-3.88_{-0.13}^{+0.13}$ & $11.59_{-0.09}^{+0.16}$ & $-1.29_{-0.07}^{+0.06}$ & $-3.92_{-0.05}^{+0.05}$ & $11.62_{-0.09}^{+0.10}$ & $7.74_{-0.06}^{+0.06}$\\
2.25 & ... & ... & ... & $-4.48_{-0.09}^{+0.07}$ & $11.58_{-0.08}^{+0.20}$ & $7.21_{-0.08}^{+0.10}$\\
3.34 & ... & ... & ... & $-5.18_{-0.13}^{+0.09}$ & $11.58_{-0.11}^{+0.26}$ & $6.51_{-0.09}^{+0.16}$\\
\hline
\end{tabular}
\label{table:params}
\end{table*}

\subsubsection{\hb~SMFs: $z$ = 0.84 - 3.24}
\begin{figure*}
\centering
\includegraphics[width=1.05\textwidth]{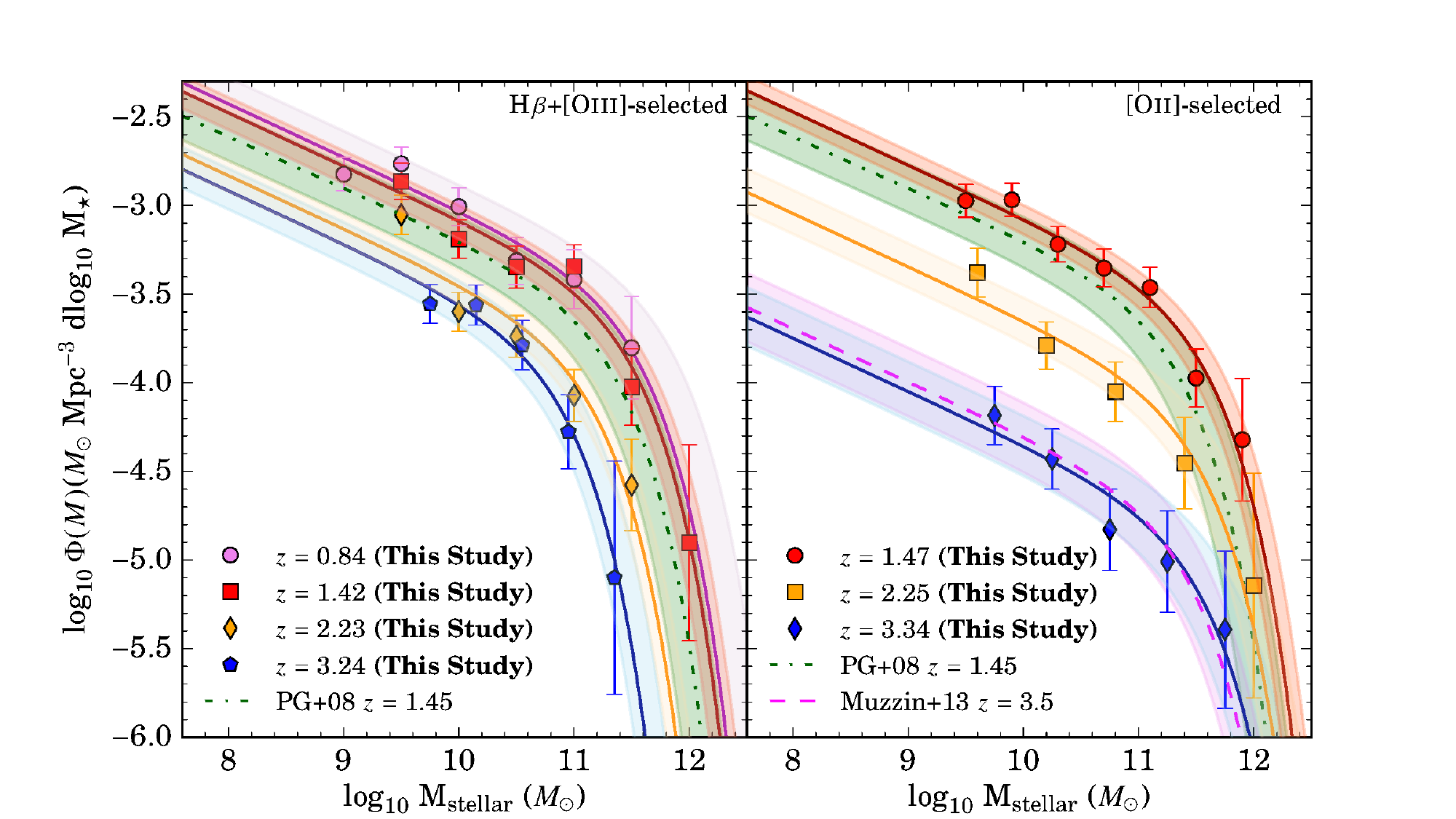}
\caption{{\it Left:} \hb~stellar mass functions and its evolution from $z \sim 0.84$ to $z \sim 3.24$ for emission-line selected sources. We find that around the $z \sim 2 - 3$, there is no significant evolution until $z < 2$. {\it Right:} \oii~stellar mass functions between $z = 1.47$ and $3.34$. For \oii, we find a strong, increasing evolution with increasing redshift in $\phi_\star$ while \hb~varies little. We also find that $M_\star$ is strongly decreasing with increasing redshift for \hb~and is relatively constant between $z \sim$ 1.47 to 3.34 for \oii.}
\label{fig:SMF}
\end{figure*}

We show on the left panel of Figure \ref{fig:SMF} the \hb~SMFs from $z = 0.84$ to 3.24 with the corresponding binned measurements and the 1$\sigma$ confidence area. The tabulated measurements are shown in Table \ref{table:HB_SMF}. We find a strong evolution in $M_\star$ where the characteristic mass increases from $z = 3.24$ to $1.42$ and then varies slowly by $z = 0.84$. This is also accompanied by an evolution in $\phi_\star$ where the normalization increases from $z = 3.24$ to $1.42$ and, just like $M_\star$, changes very little to $z = 0.84$. From the viewpoint of the cosmic SFR evolution, we are most likely seeing the rapid build-up of stellar mass between $z = 3.24$ and $z = 1.42$, followed by the decrease in stellar mass growth by $z = 0.84$ as star-formation activity in galaxies declines. We note that this could also be caused by the \hb~selection picking up different populations across cosmic time, particularly due to the change in the typical ionization parameter (see \S \ref{sec:ion}).

We compare our results with the $UVJ$-selected SF SMFs of \citet{Muzzin2013}, $NUVrJ$-selected SF SMFs of \citet{Ilbert2013}, and {\it Spitzer} IRAC selected SF SMFs of \citet{Perez-Gonzalez2008}. Not surprisingly (due to different selection), we find that our measurements, in terms of $\phi_\star$ and $M_\star$, are in disagreement with those from the literature. The only exception is the $z = 1.45$ measurement of \citet{Perez-Gonzalez2008}, which is in agreement within 1$\sigma$ of our $z = 1.42$ measurement. As stated above, we fixed $\alpha = -1.3$ based on the faint-end slope measurements from the studies mentioned above. The discrepancy is most likely based on sample selection as our sample is narrow-band selected and will select different population types in comparison to attempts at mass-selected samples such as \citet{Perez-Gonzalez2008}, \citet{Ilbert2013}, or \citet{Muzzin2013}.

We also compared our measurements to the HiZELS H$\alpha$ SMFs of \citet{Sobral2014}. We find that there is still discrepancies between our $\phi_\star$ and $M_\star$ and those of \citet{Sobral2014}. For the overlapping $z = 0.84$, 1.42, and 2.23 samples, we find disagreements in both $\phi_\star$ and $M_\star$. This discrepancy can be attributed to population differences since the \ha~samples of \citet{Sobral2014} cover the full range of star-forming galaxies (see \citealt{Oteo2015}). The issue could be that our \hb~samples (especially at higher redshifts) are missing the dustier, starburst galaxies as shown in Figure \ref{fig:UVJ} where we find that the rest-frame $UVJ$ colours are bluer with increasing redshift.

\subsubsection{\oii~SMFs: $z$ = 1.47 - 3.34}
Figure \ref{fig:SMF} presents the \oii~SMFs from $z = 1.47$ to 3.34 with the highlighted regions showing the 1$\sigma$ confidence area. The tabulated measurements are shown in Table \ref{table:OII_SMF}. We find that there is a strong evolution in $\phi_\star$ and a constant $M_\star$ for all three redshifts sampled. The quick increase in the SMFs as shown in Figure \ref{fig:SMF} could be evidence of the build-up of stellar masses due to an increase in star-forming activity towards the peak of cosmic star-formation. 

In comparison to the measurements from the literature, we find that we are in agreement with the UltraVISTA/COSMOS measurements of \citet{Muzzin2013} where they measure a $z = 3.5$ SMF with $\phi_\star = 10^{-5.10\pm0.11}$ Mpc$^{-3}$ and $M_\star = 10^{11.47\pm0.07}$ \msol~in comparison to our $z = 3.34$ SMF with $\phi_\star = 10^{-5.19^{+0.09}_{-0.13}}$ Mpc$^{-3}$ and $M_\star = 10^{11.58^{+0.26}_{-0.11}}$ \msol~(within 1$\sigma$ agreement). Note that \citet{Muzzin2013} fixed $\alpha = -1.3$ (the same that we used in fitting the SMFs). We also find agreement with the {\it Spitzer} IRAC-selected, star-forming $z = 1.45$ SMF of \citet{Perez-Gonzalez2008} where they find a $\phi_\star = 10^{-3.96\pm0.09}$ Mpc$^{-3}$ and $M_\star = 10^{11.40\pm0.10}$ \msol~with $\alpha  = -1.29\pm0.08$ (corrected from Salpeter to Chabrier IMF) in comparison to our $z = 1.47$ SMF with $\phi_\star = 10^{-3.92\pm0.05}$ Mpc$^{-3}$ and $M_\star = 10^{11.62^{+0.10}_{-0.09}}$ \msol. 

We also compare to the HiZELS H$\alpha$ SMF of \citet{Sobral2014} to the overlapping $z = 1.47$ and $z = 2.25$ measurements. As in the \hb-\ha~comparison, we find discrepancies when comparing $\phi_\star$ and $M_\star$. This discrepancy most likely arises from the \ha~sample tracing the full star-forming population, while the \oii~sample could include potential LINERs (low \oiii/H$\beta$ ratios equates to lower \oii~luminosities) and bright emitters as potential AGNs. Despite this contamination, \oii~has been shown to be a reliable star-forming indicator\footnotemark (e.g., \citealt{Hayashi2015}) and to test whether LINERs and AGNs may be contributing to this discrepancy will require spectroscopic follow-up.

\footnotetext{This is still a matter of debate as the \oii~line is also metallicity dependent (e.g., \citealt{Kewley2004}). A recent study by \citet{Darvish2015} used a sample of 58 spectroscopically-confirmed $z \sim 0.53$ star-forming galaxies and found that the dust- and metallicity-corrected SFR(\oii) was consistent up to $\sim 0.02$ dex with SFR(H$\beta$). Future spectroscopic measurements of $z > 1$ are needed to reliably ascertain the nature of \oii~as a star-formation indicator.}

\subsection{Evolution of Stellar Mass Densities}
We infer the stellar mass densities (SMDs) by integrating the stellar mass functions for the full mass range:
\begin{equation}
\rho_\star = \int_0^\infty M \Phi(M) \mathrm{d}M = \phi_\star M_\star \Gamma(2+\alpha)
\end{equation}
where $\rho_\star$ is the stellar mass density, $\phi_\star$ is the normalization, $M_\star$ is the characteristic stellar mass, and $\alpha$ is the faint-end slope. We report the SMDs in Table \ref{table:params} for all of our samples.

\begin{figure}
\centering
\includegraphics[width=1.1\columnwidth]{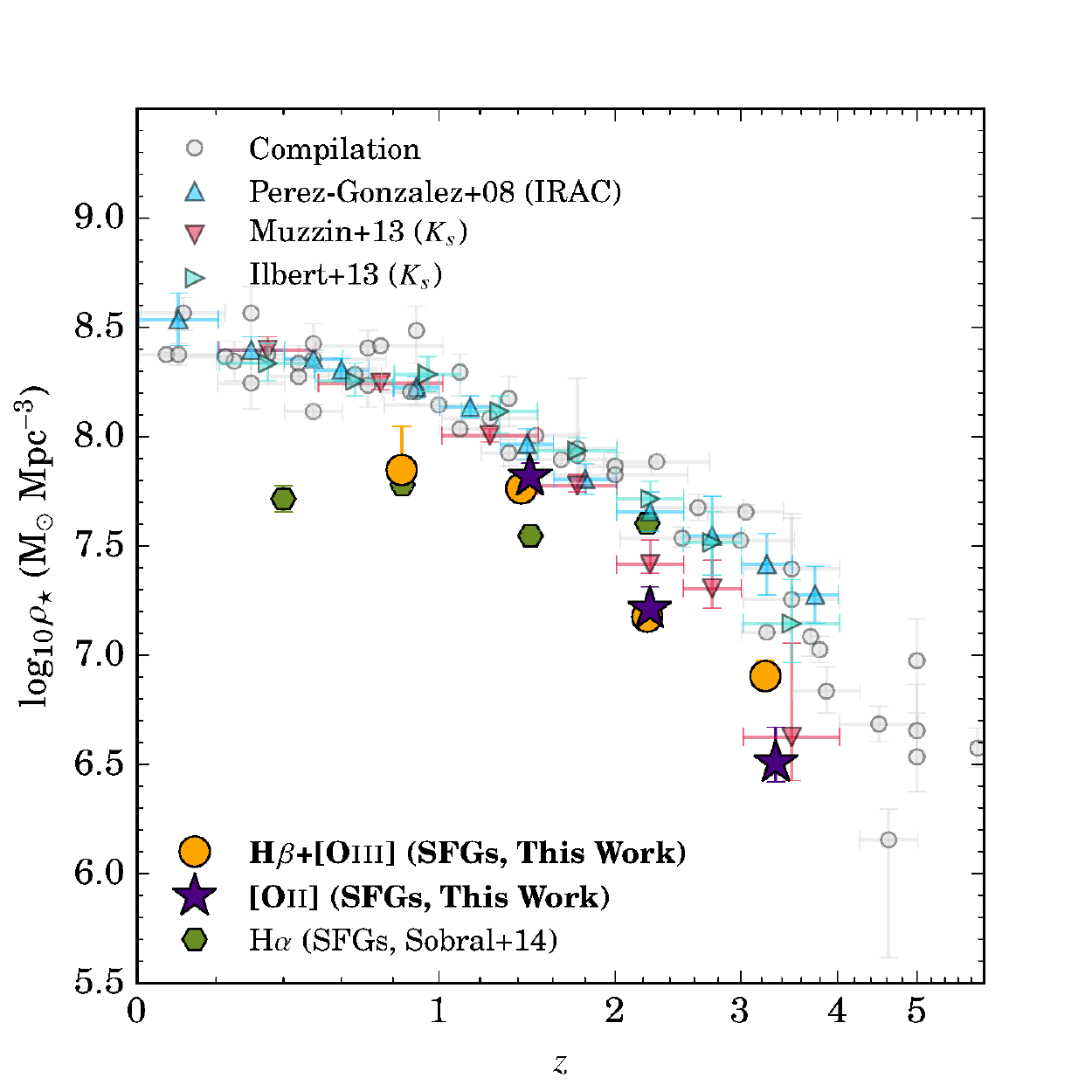}
\caption{The cosmic stellar mass density evolution of our \hb~and \oii~sample. Overlaid are the full population (star-forming + quiescent) measurements from the \citet{Madau2014} compilation. We also highlight the \citet{Perez-Gonzalez2008}, \citet{Ilbert2013}, and \citet{Muzzin2013} SMD measurements. We find that our measurements match the general picture of a fast stellar mass build-up from $z \sim 3.3$ to $z \sim 1$. By $z \sim 1$, we find that our measurements diverge from the full population literature measurements, implying that star-formation activity in emission-line selected galaxies is decreasing resulting in a slower stellar mass assembly growth and a population transition from star-forming/active to quiescent/passive systems.}
\label{fig:smd}
\end{figure}

Our measurements are shown in Figure \ref{fig:smd} for both \hb~and \oii~up to $z \sim 3.3$. We find that for $z \sim 3.3$ to $z \sim 1.5$, both samples of line emitters shown an increase in stellar mass build-up. This is consistent with the view that galaxies were producing stars at an increasing rate up to $z \sim 2$. In the case of \oii, our assessment of the SMD ends here as we have no $z < 1.5$ \oii~measurements. 

We also find that our \hb~and \oii~measurements at $z \sim 1.5$ and $\sim 2.2$, respectively, are in good agreement. For the $z \sim 3.3$ measurements, we find a discrepancy between the \hb~and \oii~measurements where the separation is $\sim 0.4$ dex. This discrepancy could be attributed to a sample bias due to the different $L_\star(z)$ cuts that were applied ($0.4 L_\star(z)$ and $0.85 L_\star(z)$ for \hb~and \oii, respectively) or even number statistics (since the \oii~$z = 3.34$ is the smallest sample being comprised of only 35 emitters, while the \hb~sample contains 179 emitters).

Figure \ref{fig:smd} also presents a comparison to the full population (star-forming + quiescent) literature compilation of \citet{Madau2014}. We also highlight the IRAC-selected full sample of \cite{Perez-Gonzalez2008}, and the COSMOS/UltraVISTA $K_s$-band measurements of \citet{Ilbert2013} and \citet{Muzzin2013}. We note that these samples have measurements for the star-forming population, although these mass-selected samples are divided by using a colour-colour selection(s) (e.g., $UVJ$) to separate the quiescent and star-forming populations. We instead use the full population literature measurements as a way to qualitatively gauge the evolution of the star-forming fraction of galaxies.

Also shown on Figure \ref{fig:smd} are the SMD measurements of the HiZELS H$\alpha$ sample from \citet{Sobral2014}. We find that our measurements are consistent with the literature in the sense that all our measurements are implying a stellar mass build-up all the way to $z = 0.84$. In comparison to the SMD compilation of \citet{Madau2014} and the measurements of \citet{Perez-Gonzalez2008}, \citet{Ilbert2013}, and \citet{Muzzin2013}, we find that our \hb~and \oii~SMDs are all below the literature, which is expected as these are for a subset (``active" galaxies) of the total population of galaxies. From $z \sim 3$ to $\sim 1.5$, this gap diminishes implying that the star-formation fraction increases up to $z \sim1.5$ where it then decreases until $z \sim 0.8$ as the gap increases. In comparison to the HiZELS H$\alpha$ measurements of \citet{Sobral2014}, we find that we are in agreement for the $z \sim 0.84$ \hb~sample. At all other redshifts, we are not in agreement, but this is due to sample biases where at $z = 1.47$ our \hb~and \oii~samples probe $\sim 0.30$ and $\sim 0.17$ dex deeper in line luminosity, respectively, than the H$\alpha$ measurements. For the $z = 2.23$ measurements, our \hb~and \oii~measurements are at the same line luminosity depth as the \ha~measurements of \citet{Sobral2014}. The inconsistency could then be attributed to the evolution of the emission lines itself.

We note that this evolution (especially at higher redshifts) could be a byproduct of the change in the physical conditions that produce these lines. Therefore, it is important to keep in mind when interpreting the results shown in Figure \ref{fig:smd} that other variables (e.g., electron densities, ionization parameter, gas abundances, metal absorption, etc.) can affect and/or drive the evolution (e.g., \citealt{Nakajima2014, Hayashi2015}). With this in mind, it becomes apparent that we must study the physical conditions of the ISM for which these lines originate from. We do this in the following sections by investigating the \ewr~evolution for each emission line, as well as the observational proxy of the ionization parameter (\ioni) and its evolution over cosmic time.

\subsection{Equivalent Widths of \hb~and \oii~Emitters}

\begin{figure*}
\centering
\includegraphics[width=0.9\textwidth]{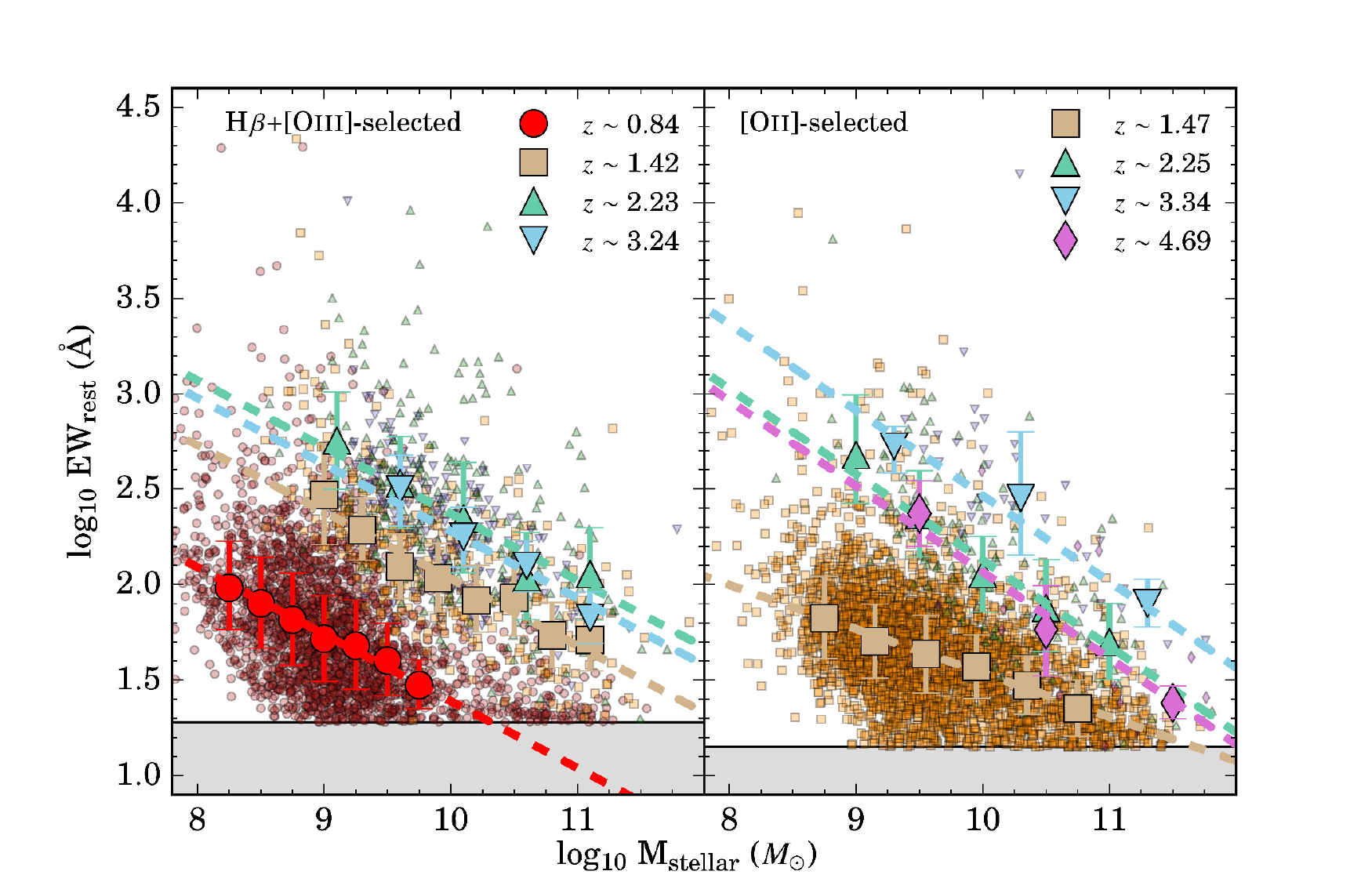}
\caption{Shown is a scatter plot of the \ewr~versus \mst~for all our samples. We also include, as larger symbols, the median \ewr~for given stellar mass bins. Highlighted in {\it grey} is the \ewr~limit, which results in an incompleteness in our sample for the high-mass sources. This effect is only seen in the \hb~$z = 0.84$ sample and to some extent in the \oii~$z = 1.47$ sample. For the other high-$z$ samples, the \ewr~limit does not cause any incompleteness in the high-mass end as we do not probe high enough masses (low \ewr) for which it must be considered.}
\label{fig:EW_Mstar}
\end{figure*}

\subsubsection{Equivalent Width -- \mst~Relation}
\label{sec:EW_Measure}
\citet{Fumagalli2012} and \citet{Sobral2014} have both shown a power-law relationship between the median \ewr(H$\alpha$) and \mst, as well as an increasing evolution in the normalization. This signifies that for every mass bin, the median \ewr~for H$\alpha$ increases with redshift. We extend this analysis for our \hb~and \oii~sample and measure the \ewr-\mst~relationship up to $z \sim 5$.

The \ewr~are calculated using Equation \ref{eqn:ewr}. Note that no dust correction has been applied to the line and continuum fluxes as we assume that $E(B - V)_{\mathrm{nebular}} \sim E(B-V)_{\mathrm{stellar}}$. The credibility of assuming that the reddening of the nebular is equivalent to that of the stellar continuum is still in debate. \citet{Calzetti2000} finds $E(B - V)_{\mathrm{nebular}} = 2.27 E(B-V)_{\mathrm{stellar}}$ for which other studies have reached the same conclusion (e.g., \citealt{Forster2009,Wild2011,Wuyts2011,Hemmati2015}). \citet{Kashino2013} measured $E(B - V)_{\mathrm{nebular}} = 1.20 E(B-V)_{\mathrm{stellar}}$ using a sample of 271 sBzK-selected, spectroscopically confirmed H$\alpha$ sources as part of the FMOS-COSMOS survey. Using 3D-{\it HST} grism spectroscopic measurements of 79 $z \sim 1$ {\it Herschel}-selected main sequence star-forming galaxies, \citet{Puglisi2016} measured $E(B - V)_{\mathrm{nebular}} = 1.07 E(B-V)_{\mathrm{stellar}}$. Recently, \citet{Shivaei2015} used a sample of 262 spectroscopically confirmed $z \sim 2$ star-forming galaxies from the MOSDEF survey and concluded that, on average, $E(B - V)_{\mathrm{nebular}} = E(B-V)_{\mathrm{stellar}}$, although they find it to dependent on SFR. \citet{Reddy2015} came to a similar conclusion that $E(B - V)_{\mathrm{nebular}} - E(B-V)_{\mathrm{stellar}} = -0.049 + 0.079/\xi$, where $\xi = 1./(\log_{10}[\mathrm{sSFR(SED)}/\mathrm{yr}^{-1}]+10)$. Due to the conflicting measurements in the literature, we find that a change in our initial assumption would result in our \ewr~measurements systematically changing by a factor of $-0.4 k(\lambda) [E(B - V)_{\mathrm{nebular}} - E(B - V)_{\mathrm{stellar}}]$ dex, where $k(\lambda)$ is the dust attenuation curve.

Figure \ref{fig:EW_Mstar} shows the full sample with the binned measurements. Because of the significant scatter, each of the binned data points represent the median \ewr~and the $1\sigma$ errors are measured via bootstrapping to incorporate the errors due to scattering. Based on the actual data points and the binned data, we can see a linear trend such that the \ewr~is increasing with decreasing stellar mass. This is also seen in the H$\alpha$ studies of \citet{Fumagalli2012} and \citet{Sobral2014}. We also highlight in Figure \ref{fig:EW_Mstar} the \ewr~cut which was used in the initial selection of narrow-band colour excess \citep{Sobral2013}. For the \hb~$z > 1$ and \oii~$z > 2$ samples, this selection does not have an effect on the medians calculated since their \ewr~are much higher than the \ewr~selection limit. Although, the line flux-limit is more important for our high-$z$ samples as the effect would be the lack of fainter emission-line sources which consequentially leads to sources with lower \ewr. For our \hb~$z = 0.84$ and \oii~$z = 1.47$ samples, the \ewr~limit affects the median \ewr~measured beyond a set mass range. We then only show median \ewr~measurements below 10$^{10}$ \msol~for \hb~and 10$^{11}$ \msol~for \oii.

\begin{table}
\centering
\resizebox{\columnwidth}{!}{ 
\begin{tabular}{lcccc}
\hline
\multicolumn{5}{c}{Parameters of the Power-Law \ewr$\propto M^\beta$}\\
\hline
$z$ & Emitter & $\beta$ & $\log_{10}$ Norm. & $\log_{10}$ Norm. ($\beta$ fixed)\\
\hline
0.84 & \hb &$-0.33\pm0.02$&$4.72\pm0.14$&$4.89\pm0.01$\\
1.42 & \hb &$-0.33\pm0.03$&$5.33\pm0.32$&$5.53\pm0.02$\\
2.23 & \hb &$-0.38\pm0.06$&$6.20\pm0.61$&$5.87\pm0.04$\\
3.24 & \hb &$-0.43\pm0.04$&$6.66\pm0.38$&$5.78\pm0.03$\\
1.47 & \oii &$-0.23\pm0.01$&$3.79\pm0.12$&$3.84\pm0.01$\\
2.25 & \oii &$-0.48\pm0.04$&$6.90\pm0.44$&$6.63\pm0.03$\\
3.34 & \oii &$-0.41\pm0.04$&$6.58\pm0.45$&$6.97\pm0.04$\\
4.69 & \oii &$-0.49\pm0.04$&$6.97\pm0.44$&$6.57\pm0.03$\\
\hline
\end{tabular}}
\caption{Shown are the fitted parameters of the power-law that relates \ewr~to \mst. We run two different fits: one for which both parameters are free and the other where $\beta = -0.35$ and $-0.45$ for \hb~and \oii, respectively. This is to ensure compatibility between samples and mitigation of the bias from selection effects when looking at the evolution of the normalization. The only exception is the $z = 1.47$ \oii, which is fitted for a constant $\beta = -0.23$ as this better fits the data.}
\label{table:powerlaw}
\end{table}

As in \citet{Fumagalli2012} and \citet{Sobral2013}, we find that the median \ewr-\mst~relationship is best fitted with a power-law of the form \ewr~$\propto M^\beta$, where $M$ is the stellar mass and $\beta$ is the power-law slope. Table \ref{table:powerlaw} shows the fitted parameters for each sample. We notice that for all \hb~samples, $\beta \sim -0.35$ which is somewhat higher than the $\beta = -0.25\pm0.01$ measured by \citet{Sobral2014} for their \ha~samples. This is also consistent with the 3D-{\it HST} $1.1 < z < 1.5$ $\beta = -0.38$ of \citet{Fumagalli2012}. The normalization is found to increase with increasing redshift and flatten out by $z = 3.24$. For the \oii~samples, we find that the $z = 1.47$ is consistent with $\beta = -0.23\pm0.01$ while the $z > 1.5$ samples have $\beta \sim -0.45$. This is consistent with the $z = 0.53$ spectroscopic \oii~measurement of \citet{Darvish2015} where they find $\beta = -0.47\pm0.06$ . We find the normalization increases up to $z = 3.34$ then seems to drop by $z = 4.69$. 

We note that this evolution is affected by systematic effects arising from selection biases. Since our sample is both \ewr-limited and luminosity-limited, we then miss lower-mass sources ($M < 10^{8.5}$ \msol) due to the luminosity-limit, and higher-mass sources ($M > 10^{10}$ \msol; for $z = 0.84$ \hb) due to the \ewr~cut at a fixed SFR. To test how the selection effects can affect our results, we use our most populated and deep samples (\hb~$z = 0.84$ and \oii~$z = 1.47$) and apply luminosity limits between 10$^{40.4}$ to $10^{41.7}$ erg s$^{-1}$ in increments of 0.1 dex and fit the same power-law to the sample. We then look at the variations in $\beta$ and the normalization as a function of the luminosity limit. We find that as the luminosity limit increases, $\beta$ becomes steeper while the normalization increases. This is expected since the two are not independent from each other. As the luminosity limit increases, then more sources with low-mass will be removed such that the median \ewr~increases more towards lower masses, resulting in $\beta$ becoming steeper and the normalization increasing.

Because of this degeneracy, we then repeat the same methodology with $\beta$ fixed to $-0.35$ and $-0.45$ for all \hb~and \oii~samples, respectively, (except for the \oii~$z = 1.47$ where $\beta = -0.23$) and fit for the normalization as a function of the luminosity limit. We find that the normalization does not change more than $< 0.1$ dex for \hb~and $< 0.01$ dex for \oii.  

The fit is shown in Table \ref{table:powerlaw} and Figure \ref{fig:EW_Mstar}. We find that the normalization evolution is in fact real and implies that with increasing redshift, the median \ewr~for a given stellar mass increases up to $z = 2.23$ for \hb~and for our \oii~sample up to $z = 3.34$.

\begin{figure*}
\centering
\includegraphics[width=1.1\textwidth]{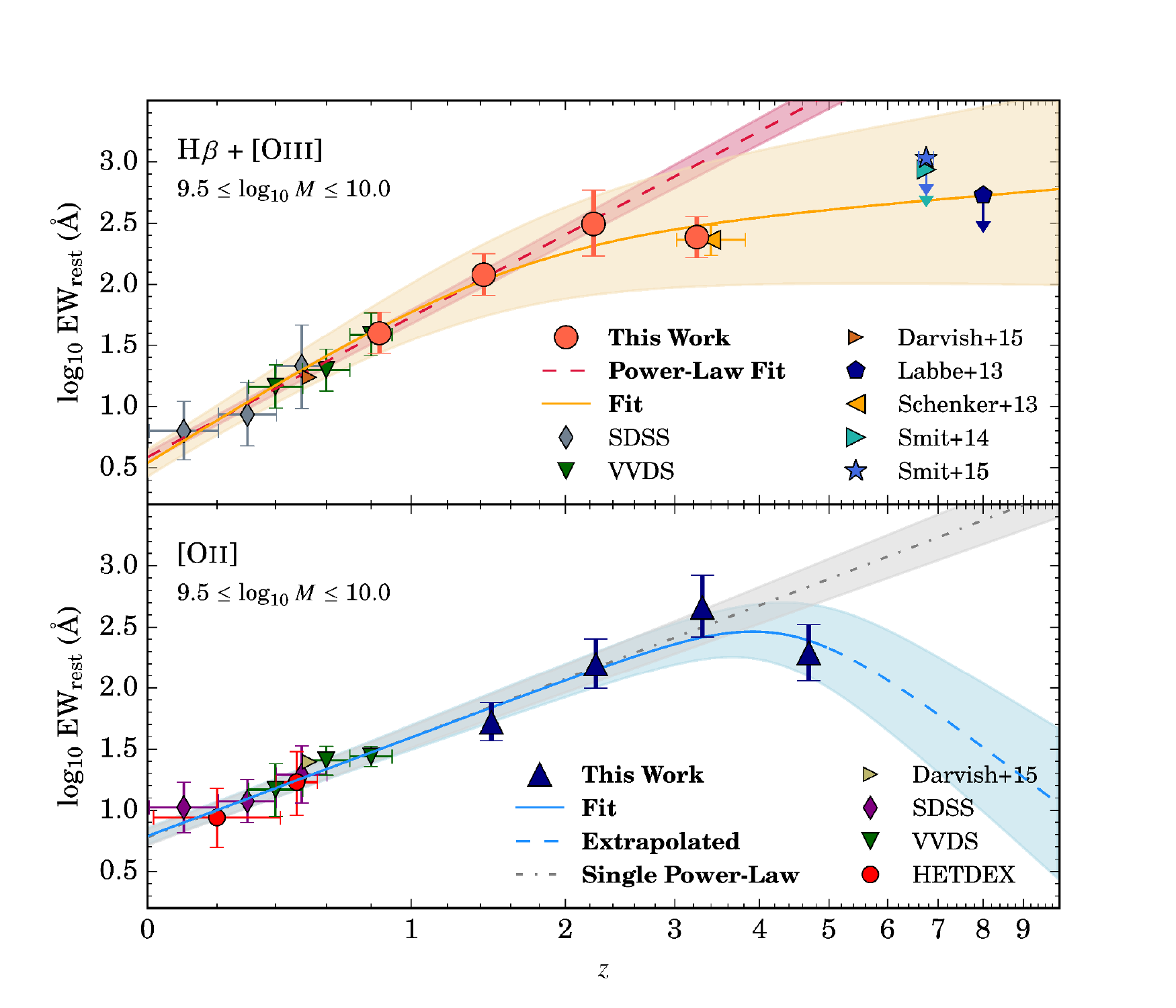}
\caption{Presented is the \ewr~evolution for sources that have $9.5 < \log{10} \textrm{\mst} < 10$ \msol. We also include measurements from the literature to constrain the low-$z$ end and to compare to our extrapolated fits in the high-$z$ regime. We fit single power-law and mixed power-law functions (combination of two power-laws) to our measurements and those from the literature. Included for each fit is the shaded 1$\sigma$ region. We find that the \ewr~evolution for \hb~flattens out to $z > 5$ and the \oii~drops in this regime. In terms of the ionization state of the gas, we find that the \ewr~evolution of both emission-lines hints to a harder ionizing source, although other factors such as metallicities and abundances can affect the evolution as well.}
\label{fig:ew_HB_OII}
\end{figure*}

\subsubsection{Evolution of Equivalent Widths with Redshift}
\label{sec:ew_evolution}
Based on the normalization seen in the \ewr-\mst~relationship, we study the evolution of the normalization and compare with measurements from the literature. Figure \ref{fig:ew_HB_OII} shows the evolution of the median \ewr~for our \hb~and \oii~measurements. For each measurement, we make a correction for the skewness of the mass distribution per each measurement. Since we select a specific mass range between $10^{9.5} < \log_{10} M < 10^{10.0}$ \msol\footnotemark, we ideally would want the median stellar mass of each of our measurements to be equal to $10^{9.75}$ \msol. This is not always the case such that the stellar mass distribution is skewed from a normal distribution. Because of the dependency between \ewr~and \mst, not correcting for the skewness in the distribution would result in systematic increases/decreases in the measured median \ewr~(corresponding to the mean stellar mass measured). To correct for this, we measure the mean mass for each sample and compute the inferred \ewr~from the corresponding fit. We then compute based on the fit what the median \ewr~should be at the center of the mass bin ($10^{9.75}$ \msol) and then subtract both measurements to get a correction factor. The result is that the median \ewr~increases/decreases ($\sim 0.1$ dex) based on whether the mean mass was above/below $10^{9.75}$ \msol.

\footnotetext{We select this mass range to be consistent with the $z > 5$ studies (e.g, \citealt{Labbe2013, Smit2014, Rasappu2015, Smit2015}) and also because it corresponds to the peaks in our stellar mass distributions as shown in Figure \ref{fig:hist}.}

Included in Figure \ref{fig:ew_HB_OII} are the \hb~measurements from the literature \citep{Labbe2013,Schenker2013,Smit2014,Smit2015}. To ensure a constrained \ewr($z = 0$), we compute the median \ewr~from the SDSS-III/BOSS-DR12 spectroscopic sample \citet{Thomas2013} by selecting only emission-lines with \ewr$> 3$ \AA~to ensure that the measured \ewr~is not dominated by uncertainties in the stellar continuum subtraction \citep{Fumagalli2012} and all galaxies that were classified as star-forming based on the BPT diagram. The VVDS catalog of \citet{Lamareille2009} was also included where only galaxies identified as star-forming were selected. We also include the \oiii~$z \sim 0.53$ \ewr~measurements from the Keck DEIMOS spectroscopic sample of \citet{Darvish2015}.

\begin{figure}
\centering
\includegraphics[width=1.1\columnwidth]{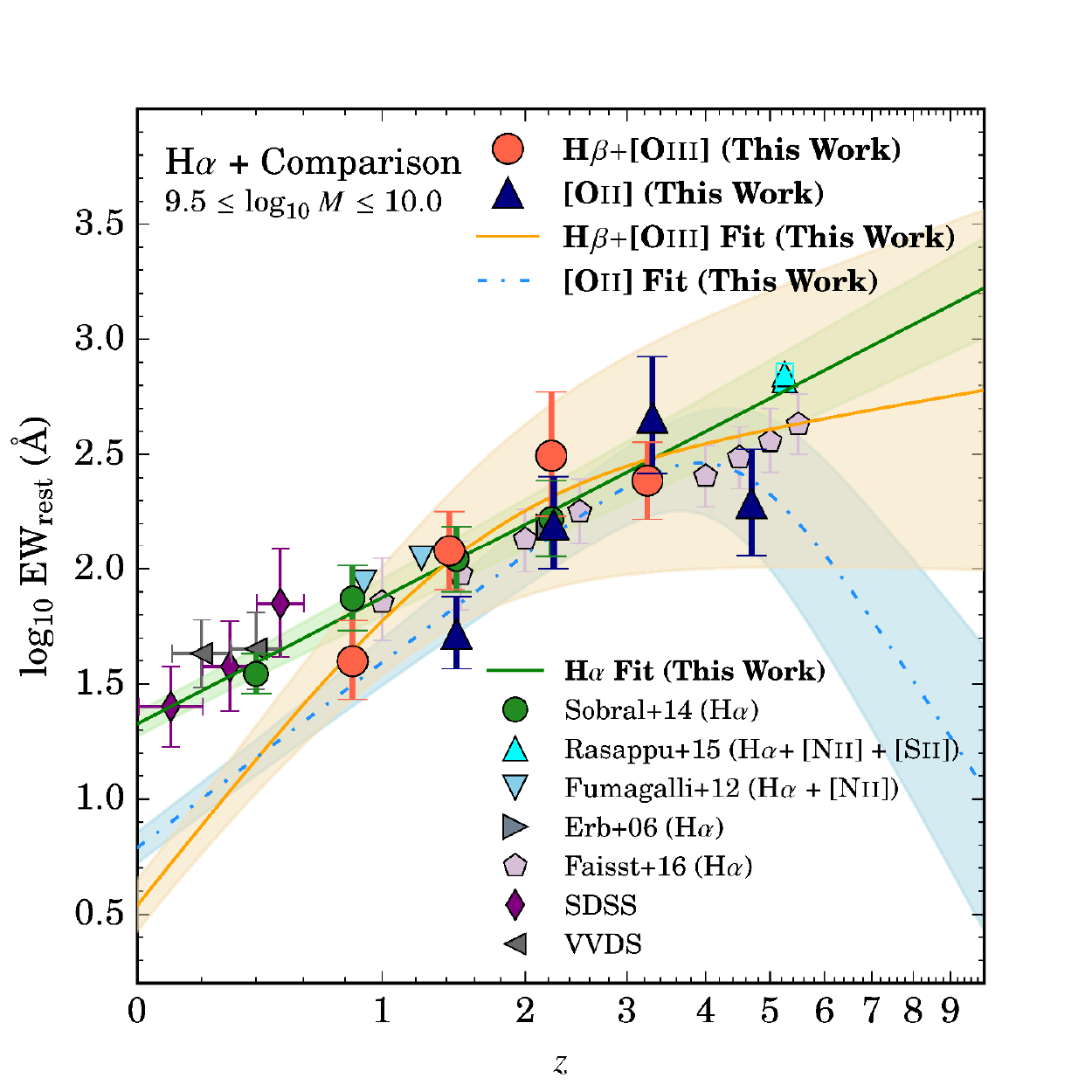}
\caption{The \ewr~evolution of major rest-frame optical emission lines within $9.5 < \log_{10} \textrm{\mst} < 10$ \msol. We include our empirical fits of the \hb~and \oii~\ewr~evolution in order to study how the \ewr~evolves per nebular emission-line. We find that the \hb~\ewr~drops faster from high-$z$ to low-$z$ than the other emission-lines. This is followed by \oii~and then by \ha~such that the \hb~\ewr~at $z = 0$ is weaker compared to \oii, which is also weaker than \ha. The drops are in order of higher to lower ionization potentials such that in the low-$z$ Universe, higher ionization potential lines have lower \ewr~relative to low ionization potential lines. We also find that the \hb~\ewr~is much higher than \oii~for $z > 5$, implying a Universe with extreme ionizing sources that easily can produce the \oiii~line.}
\label{fig:ew_evolution}
\end{figure}

For the \oii~sample, we also compute the median \ewr~from the HETDEX survey \citep{Adams2011, Bridge2015} and remove any sources with X-ray detection found by \citet{Bridge2015} to eliminate AGN contamination. We also include the \oii~$z \sim 0.53$ \ewr~measurements from \citet{Darvish2015}. Figure \ref{fig:ew_evolution} shows the H$\alpha$ \ewr~evolution found in the literature \citep{Erb2006, Fumagalli2012, Sobral2014, Rasappu2015,Faisst2016} in comparison to the \ewr~evolution of the \hb~and \oii~samples. We selected a mass range of $10^{9.5} < M < 10^{10.0}$ \msol~for all determinations of the \ewr~evolution. Changing the mass range used in Figures \ref{fig:ew_HB_OII} and \ref{fig:ew_evolution} only changes the normalization because of the power law relationship shown in Figure \ref{fig:EW_Mstar}. Furthermore, all errors presented in Figure \ref{fig:ew_HB_OII} and \ref{fig:ew_evolution} for our sample and the SDSS, VVDS, and HETDEX determined measurements are based on a bootstrapping assessment to calculate the 95\% confidence intervals.

To ensure that all the literature data is consistent and comparable with our data set, we correct the literature measurements to match our IMF (convert from the literature-assumed IMF to \citet{Chabrier2003} IMF) and also cover the same mass range ($10^{9.5} < M < 10^{10.0}$ \msol). We also make another correction for the $z > 5$ \hb~literature data points \citep{Labbe2013,Smit2014,Smit2015} as described in Appendix \ref{sec:correction} to take into account the contribution of H$\beta$ in the total \ewr~measured in these studies.

We fit the evolution of the \ewr~($z$) to a mix of power-laws of the form:
\begin{equation}
\mathrm{EW}_\mathrm{rest}(z) = \mathrm{EW}_\mathrm{rest}(z = 0) \frac{(1+z)^\gamma}{1 + [(1+z)/c]^\epsilon}
\label{eqn:power-law}
\end{equation}
where $\gamma$ and $\epsilon$ are the power-law slopes. This functional form is similar to that used by \citet{Madau2014} to model the cosmic SFRD evolution. For the \hb~sample, we only use our measurements, our SDSS and VVDS determinations, and the upper limits set by \citet{Labbe2013} and \citet{Smit2014,Smit2015} to constrain the fit. For the \oii~sample we use our measurements, the SDSS and VVDS determinations, and the HETDEX measurements. The fitted parameters are shown in Table \ref{table:evol_params} for our sample of \hb~and \oii~emitters, as well as the HiZELS \ha~sample from \citet{Sobral2014}, which was further constrained by the SDSS and VVDS data. We also overlay the fits and their 1$\sigma$ error range on Figure \ref{fig:ew_HB_OII}. Note that we also fit a simple power-law of the form $(1+z)^\gamma$. This functional form has been shown to work for the \ha~\ewr~evolution (e.g., \citealt{Fumagalli2012, Sobral2014, Rasappu2015, Marmol2015}).

\begin{table}
\centering
\resizebox{\columnwidth}{!}{ 
\begin{tabular}{c c c c c c}
\hline
\multicolumn{6}{c}{Power-Law Fit Parameters}\\
\hline
Sample & Model & $\mathrm{EW}_\mathrm{rest}(z = 0)$ (\AA) & $\gamma$ & $\epsilon$ & $c$\\
\hline
\hb & Single & $3.85\pm0.34$ & $3.81\pm0.14$ & ... & ... \\
\hb & Mixed & $3.53\pm0.90$ & $4.53\pm0.63$ & $3.93\pm 0.47$ & $2.57\pm0.46$\\
\oii & Single & $6.00\pm0.90$ & $2.72\pm0.19$ & ... & ... \\
\oii & Mixed & $6.14\pm0.95$ & $2.68\pm0.25$ & $8.09\pm1.38$ & $5.35\pm0.54$\\

\ha & Single & $21.14\pm2.54$ & $1.82\pm0.20$ & ... & ... \\

\hline
\end{tabular}}
\caption{Measurements of the Power-Law Parameters. Two different models were used to fit the data. Those listed as ``single" refer to a single power-law of the form $(1 + z)^\gamma$ and those listed as ``mixed" refer to the model as defined in Equation \ref{eqn:power-law}. }
\label{table:evol_params}
\end{table}

As shown in Figure \ref{fig:ew_HB_OII}, a single power-law would match our \hb~measurements and others drawn from the literature up to $z \sim 2$. For $z > 2$, a single power-law model would pass above the upper limits set by \citet{Smit2014, Smit2015} and \citet{Labbe2013} hinting that the slope becomes shallower and deviates from a simple power-law form. Also, our $z = 3.24$ and the $z \sim 3.5$ measurement of \citet{Schenker2013} both provide evidence that the evolution becomes shallower. The change in the slope of the \ewr~evolution has also been recently detected by \citet{Marmol2015} where they use grism spectroscopy of the H$\alpha$ line from the 3D-{\it HST} survey and samples of spectroscopically confirmed and photometric-redshift selected galaxies from CANDELS within the redshift interval $1 < z < 5$. \citet{Faisst2016} also reports a change in the power-law slope with increasing redshift up to $z \sim 6$ where the power-law deviates from $(1+z)^{1.8}$ to $(1+z)^{1.3}$. We use the mixed power-law model shown in Equation \ref{eqn:power-law} to incorporate the deviation from a single power-law and fit to our measurements, the SDSS determinations, the $z \sim 3.5$ measurements of \citet{Schenker2013}, and the upper limits set by \citet{Labbe2013} and \citet{Smit2014, Smit2015}. We find that the model defined in Equation \ref{eqn:power-law} better fits the observed measurements.

The lower panel of Figure \ref{fig:ew_HB_OII} shows the \oii~\ewr~evolution up to $z \sim 5$, along with measurements from HETDEX, VVDS, SDSS, and \citet{Darvish2015}. Our measurements are the first that cover the $z \sim 1.5$ to 5 range allowing us to compare to the $z < 1$ regime. We initially fit to a single power-law and find that the \oii~evolution increases up to $z \sim 3$. There is some evidence in our measurements for a drop from $z \sim 3$ to $\sim 5$, but more measurements have to be made in the $z > 3$ regime in order to confirm the decreasing evolution. To incorporate this drop seen between our $z = 3.34$ and $z = 4.69$ measurements, we fit using the model described in Equation \ref{eqn:power-law}.

Our \hb~and \oii~fits using the mixed power-law model described in Equation \ref{eqn:power-law} are shown in Figure \ref{fig:ew_HB_OII} with the measured parameters described in Table \ref{table:evol_params}. We find an increasing evolution in the \ewr(\hb) and \ewr(\oii) up to $z \sim 2 - 3$. The \hb~evolution trend becomes shallower from $z \sim 2$ to higher $z$. This is constrained by our $z \sim 2.23$ and $\sim 3.34$ measurements, the $z \sim 3.5$ measurement of \citet{Schenker2013}, and the recent measurements of \citet{Labbe2013} and \citet{Smit2014, Smit2015}. The literature measurements can be interpreted as upper limits since they require a significant excess in the {\it Spitzer} IRAC bands to be detected. But this assumes that the UV is bright enough that the highest EW sources are detected. Based on this interpretation, we can constrain the high-$z$ end using our $z > 2$ measurements with the condition that the fit cannot exceed the upper limits. 

Beyond $z > 3$, we find some evidence that the \ewr(\oii) is decreasing to higher $z$. Currently, there are no other measurements in the literature that cover this redshift regime. Our \ewr(\oii) measurements are the first presented in the literature at these redshifts for which we can assess the cosmic evolution of the \oii~equivalent width. Future studies from the next-generation of telescopes and space observatories will better constrain the \ewr(\oii) evolution. Based on our results, we can conclude that there is some evidence of a decrease in the \ewr(\oii) for $z > 3$. It may not be surprising then that high-$z$ UV studies (e.g., \citet{Smit2014,Smit2015}) do not find strong evidence for \oii~but do find \oiii~since, based on our measurements and the mixed power-law fits, the \oii~\ewr~is significantly lower than \oiii. This could be due to a combination of changes in the oxygen abundances and ionization state of the gas.

We also show in Figure \ref{fig:ew_evolution} the comparison of the \ewr(\ha) evolution, measured from the HiZELS \ha~sample of \citet{Sobral2014}, with our \ewr(\hb) and \ewr(\oii) measured evolution. We find that based on the fits, \ewr(\hb) drops from high to low-$z$ the fastest, followed by \oii~and then by \ha. In terms of the required ionization potentials to form these lines, it is then not surprising that the \ewr(\hb) drops the fastest since it requires a harder ionizing source (photons with $\approx 35.12$ eV) to cause a strong [O{\sc iii}] line. This is then followed by \oii~($\approx 13.62$ eV) and \ha~($\approx13.60$ eV) in decreasing order of required ionization potentials. From a broader point-of-view, the \ewr~decline in order of ionization potentials matches the current view of cosmic star-formation activity which has been in decline for the $\sim 11$ Gyr. A decrease in star formation rates results in the decrease of bright, massive stars that can create UV photons to form the emission lines we observe. Although other factors, such as metallicities, densities, electron temperatures, and abundances can also play a crucial role in the decrease of the \ewr.

\subsubsection{Evolution of the Ionization State}
\label{sec:ion}

We have shown in Figures \ref{fig:ew_HB_OII} and \ref{fig:ew_evolution} the evolution in the three major nebular emission-lines associated with star-formation to high-$z$. Based on this evolution, we investigate how the \ioni~ratio changes with redshift. The \ioni~line ratio is an important observational proxy of the ionization state of the gas since the \oiii~line has a higher ionization potential compared to the \oii~line and has been used in many studies in the literature (e.g., \citealt{Nakajima2013, Nakajima2014, Hayashi2015}). We note that the \ioni~line ratio is also dependent on stellar mass and metallicity (e.g., \citealt{Kobulnicky2004,Liu2008,Hayashi2015,Kewley2015}). To properly understand the dependency of \ioni~with the ionization parameter, stellar mass, gas-phase abundances and metallicities requires spectroscopic follow-up. In this section, we present our analysis of the \ioni~evolution in terms of the evolution in the ionization parameter but caution the reader that other factors affect this evolution as well.

If we assume that our \hb~samples are primarily \oiii~emitters (see discussion in \S \ref{sec:hb_o3}), then we can take our observed \hb~and \oii~\ewr~and measure the ratio to determine \ioni. We take the ratios of the equivalent widths rather than the ratios of the emission lines as the dependency on dust correction is eliminated with the assumption that $E(B-V)_\mathrm{nebular} \sim E(B-V)_\mathrm{stellar}$ (see discussion in \S \ref{sec:EW_Measure} on how this assumption affects the results). An issue that arises is that the continuum flux at rest-frame $3727$\AA~and $5007$\AA~may not be equivalent/similar. To test how this can affect our measurements of \ioni, we compare the \ewr(\oiii)/\ewr(\oii) and the $F_{\textrm{\oiii}}/F_\textrm{\oii}$ line ratios from the SDSS-III/BOSS-DR12 \citep{Thomas2013} and VVDS \citep{Lamareille2009} catalogs. This comparison is shown in Appendix \ref{sec:SDSS_VVDS}. We find that using the \ewr~to measure \ioni~is consistent, on average, with using the line fluxes with a negligible systematic offset arising from the differing continuum fluxes ($-0.06$ and $-0.04$ dex for SDSS and VVDS, respectively; see Figure \ref{fig:EW_FLUX_comp}).

\begin{figure}
\centering
\includegraphics[width=1.1\columnwidth]{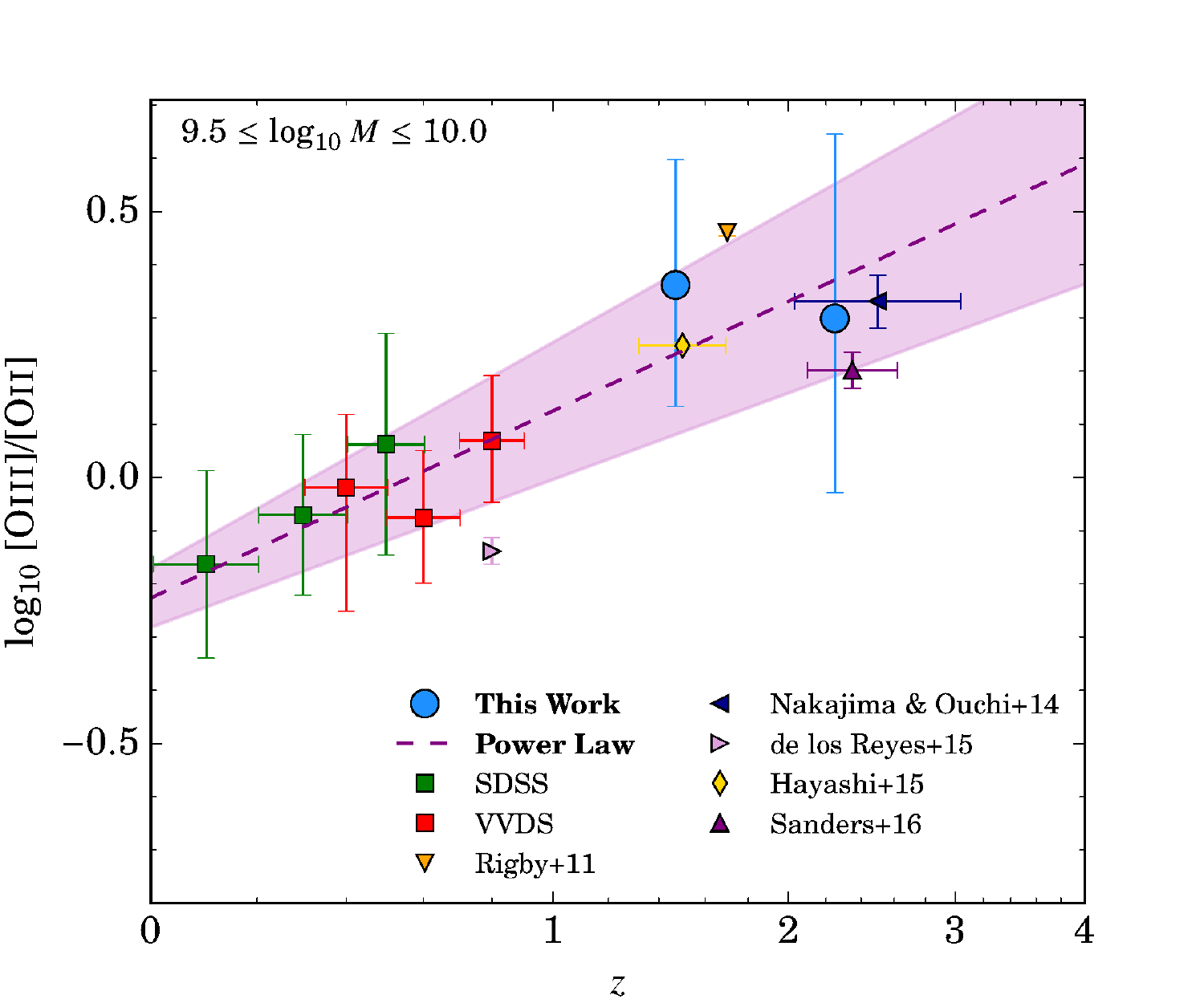}
\caption{Shown is the \ioni~evolution. Included are measurements from SSDS and VVDS, as well as other spectroscopic measurements from the literature. We find that the \ioni~increases with redshift suggesting a harder ionizing source at higher redshifts. Although, we note that evolution in the metallicity and abundances can also influence the \ioni~evolution. In terms of the ionizing source, the \ioni~evolution shown here explains why recent studies have detected emission lines that require high ionization potentials at $z > 6$ (e.g., \citealt{Vanzella2010,Sobral2015,Stark2015a,Stark2015b}).}
\label{fig:ionization}
\end{figure}

Figure \ref{fig:ionization} shows the \ioni~evolution with our observational measurements at $z = 1.47$ and 2.25 along with measurements we computed from SDSS-III/BOSS-DR12 \citep{Thomas2013} and VVDS \citep{Lamareille2009,LeFevre2013}. We also include the $z \sim 2.3$ measurement of MOSDEF \citep{Sanders2016}\footnotemark, the $z \sim 1.5$ measurement of \citet{Hayashi2015}, and the $z \sim 1.7$ measurement of \citet{Rigby2011}. We exclude our $z = 3.34$ measurement due to the \oii~sample size (13 sources) being $\sim 5$ times smaller in comparison to the \hb~sample size (62 sources) which would make the two samples incomparable. Overall, our measurements combined with those from the literature show that the \ioni~ratio is increasing up to $z \sim 3$ such that at higher redshifts the ionizing source was much harder. When we compare our measurements with those within the same redshift range, we find that we are within $1\sigma$ agreement. We note that the literature measurements are for the mass range $9.5 < \log_{10} M < 10.0$ \msol.

\footnotetext{The \citet{Sanders2016} measurement was recomputed to only cover the $9.5 < \log_{10} M < 10.0$ \msol~since \ioni~is also a function of stellar mass (e.g., \citealt{Hayashi2015}). We recompute their \ioni~measurement and calculate the errors via bootstrapping. The measurement cited in \citet{Sanders2016} is $\log_{10}\textrm{\ioni} = 0.10^{+0.37}_{-0.11}$.}

We fit the evolution of the \ioni~ratio to a power-law of the form:
\begin{equation}
\textrm{\ioni} = \textrm{\ioni}(z = 0) (1 + z)^\eta
\end{equation}
where we find \ioni$(z = 0) = 0.59\pm0.07$ (normalization) and $\eta = 1.17\pm0.24$ (power-law slope). We only use the SDSS, VVDS, and our measurements to fit for the power-law. The fit along with the 1$\sigma$ region is shown in Figure \ref{fig:ionization} and matches well with the observed data points not used in the fitting process. Based on our power-law model, the \ioni~ratio is predicted to continue to increase with redshift. This matches with the fits shown on Figure \ref{fig:ew_HB_OII} where we find that the evolution of \ewr(\hb) becomes shallower and the \ewr(\oii) drops significantly. The situation could be that the hardness of the ionizing source increases when going back in cosmic time such that the production of an \oii~emission-line is suppressed as electrons in doubly-ionized oxygen are unable to transition to lower energy levels when bombarded by highly energetic photons and free electrons. We note that this can also be the byproduct of changing metallicities and abundances. The physical source of this is still in debate, but lower amounts of metal coolants and dust, higher star formation activity and efficiency, and even changes in the initial mass function can influence the hardness of the ionizing source.

Our results for the \ioni~evolution and its extrapolation to $z > 3$ can also explain why recent spectroscopic observations are able to find emission-lines from high ionization potential transitions (e.g., C{\sc iii}], C{\sc iv}, N{\sc iv}, He{\sc ii}). \citet{Stark2014} spectroscopically observed 17 $z \sim 2$ gravitationally-lensed galaxies to find strong N{\sc iv}], O{\sc iii}], C{\sc iv}, Si{\sc iii}], and C{\sc iii}] emission-lines requiring photons with energies $> 47$ eV, much higher than the local Universe. Their argument is using such emission-lines that require high ionization energies could be used in conjunction with Ly$\alpha$ to study reionization. This led to the spectroscopic detection of N{\sc iv} ($z = 5.56$; \citealt{Vanzella2010}), C{\sc iii}] ($z \sim 6 - 7$; \citealt{Stark2015a}), and C{\sc iv} ($z = 7.045$; \citealt{Stark2015b}) emitters, such that the ionizing source is much harder with increasing redshift. An even more extreme case is the recent discovery of He{\sc ii} in the COSMOS Redshift 7 (CR7) source \citep{Sobral2015b}. To produce this emission line requires ionizing photons with energy $\sim 54$ eV and has been attributed to the presence of Pop{\sc iii} stars or direct collapse black holes (e.g., \citealt{Pallottini2015,Dijkstra2016,Visbal2016}). The following studies comprise a handful of sources but match our extrapolation of the \ioni~evolution to show that the ionization potential increases with redshift. Future studies using the next-generation space-based observatories (e.g., $JWST$) could spectroscopically observe the traditional optical emission-lines for $z > 5$ (falls in observer-frame infrared) and assess the ionization state of the gas with better accuracy. For now, we present our extrapolated $z > 3$ results as a prediction that can be tested by future high-$z$ studies.

\section{Conclusions}
\label{sec:conclusion}

We have presented the evolution of the stellar mass functions and densities up to $z \sim 3$, the evolution of the rest-frame equivalent widths up to $z \sim 5$, and the evolution of the ionization parameter as described by the \ioni~ratio up to $z \sim 3$. The main results of this study are the following:
\begin{enumerate}[label=(\roman*),leftmargin=0.5cm]
\item In conjunction with the widely used $UVJ$ colour-colour classification scheme, we find that $\sim 98\%$ of all \hb~and \oii~emitters are classified as ``active" (star-forming or AGN) galaxies. 

\item The stellar mass functions of \hb~emitters show a strong, increasing evolution in $M_\star$ from $10^{10.96^{+0.15}_{-0.08}}$ \msol~to $10^{11.60^{+0.29}_{-0.13}}$ \msol~and a weak, decreasing evolution in $\phi_\star$ from $10^{-3.87^{+0.06}_{-0.11}}$ Mpc$^{-3}$ to $10^{-4.16\pm0.08}$ Mpc$^{-3}$ with increasing redshift. The opposite trends are seen for the stellar mass functions of \oii~emitters from $z = 1.47$ to $z = 3.34$ where an unchanging $M_\star \sim 10^{11.60}$ \msol~is seen for all redshifts sampled and a strong, increasing evolution in $\phi_\star$ from $10^{-5.18^{+0.09}_{-0.13}}$ Mpc$^{-3}$ to $10^{-3.92\pm0.05}$ Mpc$^{-3}$ with decreasing redshift. 

\item The similarity between the $z = 0.84$ and $1.42$ \hb~SMFs and the rise in the SMFs between $z = 3.24$ to $z = 1.42$ is probable evidence for the rapid stellar mass build-up followed by its decay due to the decrease of star-formation activity in the Universe. The stellar mass functions of the \oii~emitters all shows rapid build-up of stellar masses from $z = 3.34$ to $z = 1.47$ for \oii-selected galaxies.

\item Stellar mass densities of our \hb~and \oii~emitters, in conjunction with the HiZELS H$\alpha$ SMDs of \citet{Sobral2014}, show how the evolution in the SMDs traces that of the full sample (passive + active) as found in the literature. By $z = 0.84$, we find that the SMDs deviate from the full population implying the transition of active galaxies into passive galaxies. This ties into the picture of decreasing star-formation activity in the Universe.

\item The relationship between \ewr~and stellar mass for \hb~and \oii~emitters up to $z \sim 3$ and $\sim 5$, respectively, is studied for the first time where we find a power-law relationship between the two physical properties as seen in H$\alpha$ studies (e.g., \citealt{Fumagalli2012,Sobral2014}). We find that all our \hb~samples are best represented by $\textrm{\ewr} \propto M^{-0.35}$ and the \oii~samples as $\propto M^{-0.45}$. The $z = 1.47$ \oii~sample has a shallower trend best fit as $\propto M^{-0.23}$.

\item We find that the \hb~\ewr~increases from $z = 0$ to $z \sim 2$ by a factor of $\sim 100$. From $z \sim 2$ to $\sim 8$, we find evidence for a shallower trend by using the {\it Spitzer} IRAC measurements of \citet{Labbe2013} and \citet{Smit2014,Smit2015} as upper limits and also the deviation from the $z = 0 - 2$ power-law seen by our $z = 3.24$ \ewr~and the $z \sim 3.5$ Keck/MOSFIRE \ewr~measurement of \citet{Schenker2013}.

\item We present the first measurement of the \oii~\ewr~out to $z \sim 5$. We find that the \oii~\ewr~increases by a factor of $\sim 60$, followed by a decrease in \ewr~to higher redshift. This could be one reason why no high-$z$ measurements of \oii~exists in the $z > 5$ regime from UV studies that are finding ubiquitous high \hb~EW sources. 

\item We study the evolution of the ionization state of the gas using the \ioni~line ratio. The line ratio increases beyond $z \sim 3$ such that the higher the redshift, the harder the ionizing source. This could explain the lack of \oii~detections at high-$z$. The harder ionizing source (e.g., high energy photons coming from massive stars) could suppress the \oii~line while producing a stronger \oiii~line as the doubly-ionized oxygen atoms are bombarded with highly energetic photons and free electrons such that they can not make the transition to produce an \oii~line. The harder ionizing source can also explain the recent detections of emission lines with high ionization potentials at $z \sim 5$ to $\sim 7$ (e.g., C{\sc iii}], C{\sc iv}, He{\sc ii}, N{\sc iv}). The physical reason for a harder ionizing source is still in debate and can be explained as changes in the amount of metal coolants, dust, star-formation activity and efficiency, and possibly even a varying initial mass function.

\end{enumerate}

Our results present a clearer picture of the \ewr~of the \hb~and \oii~lines, as well as an understanding of how the strengths of these lines and its dependency on the changes in the ionization state of the gas can explain the recent developments in detecting \hb~at $z \sim 6 - 8$ and other emission lines that arise from transitions involving high ionization potentials. The results highlighted in this paper prepare for the next-generation of ground-based telescopes (e.g., Thirty Meter Telescope) and state-of-the-art space-based observatories (e.g., {\it JWST}, {\it EUCLID}, {\it WFIRST}) by presenting an outline of the evolution of the \ewr~and the \ioni~line ratio and predictions for the high-$z$ Universe that can better our understanding of the physical conditions for which forms the observed \ewr~and \ioni~line ratios.

\section*{Acknowledgments}
We thank Philip Best, Rychard Bouwens, Naveen Reddy, Irene Shivaei, and Anahita Alavi for their insightful discussions and comments. The catalogs used in this analysis are publicly available from \citet{Sobral2013}.  

DS acknowledges financial support from the Netherlands Organisation for Scientific research (NWO) through a Veni fellowship, from FCT through a FCT Investigator Starting Grant and Start-up Grant (IF/01154/2012/CP0189/CT0010) and from FCT grant PEst-OE/FIS/UI2751/2014. IRS acknowledges support from ST7C (ST/L00075X/1), the ERC Advanced Grant DUSTYGAL (321334) and a Royal Society/Wolfson Merit award. BD acknowledges financial support from NASA through the Astrophysics Data Analysis Program (ADAP), grant number NNX12AE20G. JPS gratefully acknowledges support from a Hintze Research Fellowship.

\bibliography{sm_ew_paper}

\appendix
\section{Stellar Mass Functions}

\newpage

\begin{table}
\caption{\hb-selected stellar mass function. Shown are the stellar mass bins ($\log_{10} M$), the number of sources per bin (\#), the observed ($\Phi_{\mathrm{obs}}$) and final ($\Phi_{\mathrm{final}}$) stellar mass distribution per stellar mass bin, and the comoving volume per bin. $\Phi_{\mathrm{final}}$ includes the completeness, EW, and filter profile corrections.}
\resizebox{\columnwidth}{!}{%
\begin{tabular}{c c c c c}
\hline
$\log_{10} M$ & \#  & $\Phi_{\mathrm{obs}} $ & $\Phi_{\mathrm{final}}$ & Volume \\
(M$_\odot$) & & (Mpc$^{-3} ~ \rm{d}\log_{10} \rm{M}$) & (Mpc$^{-3} ~ \rm{d}\log_{10} \rm{M}$) & ($10^5$ Mpc$^{3}$)\\
\hline
{\bf z = 0.84} & & & & \\
9.00$\pm0.25$ & $185$ & $-2.95$ & $-2.82\pm0.09$ & $3.33$\\
9.50$\pm0.25$ & $185$ & $-2.95$ & $-2.76\pm0.09$ & $3.33$\\
10.00$\pm0.25$ & $64$ & $-3.42$ & $-3.01\pm0.10$ & $3.33$\\
10.50$\pm0.25$ & $23$ & $-3.86$ & $-3.31\pm0.13$ & $3.33$\\
11.00$\pm0.25$ & $11$ & $-4.18$ & $-3.42\pm0.17$ & $3.33$\\
11.50$\pm0.25$ & $3$ & $-4.74$ & $-3.80\pm0.29$ & $3.33$\\
\hline
{\bf z = 1.42} & & & & \\
9.50$\pm0.25$ & $111$ & $-3.03$ & $-2.86\pm0.10$ & $2.37$\\
10.00$\pm0.25$ & $80$ & $-3.40$ & $-3.19\pm0.11$ & $4.06$\\
10.50$\pm0.25$ & $54$ & $-3.57$ & $-3.35\pm0.12$ & $4.06$\\
11.00$\pm0.25$ & $44$ & $-3.66$ & $-3.34\pm0.12$ & $4.06$\\
11.50$\pm0.25$ & $8$ & $-4.40$ & $-4.02\pm0.22$ & $4.06$\\
12.00$\pm0.25$ & $1$ & $-5.31$ & $-4.90\pm0.55$ & $4.06$\\
\hline
{\bf z = 2.23} & & & & \\
9.50$\pm0.25$ & $74$ & $-3.26$ & $-3.05\pm0.11$ & $2.72$\\
10.00$\pm0.25$ & $77$ & $-3.83$ & $-3.60\pm0.11$ & $10.45$\\
10.50$\pm0.25$ & $53$ & $-4.00$ & $-3.74\pm0.12$ & $10.68$\\
11.00$\pm0.25$ & $22$ & $-4.39$ & $-4.07\pm0.15$ & $10.68$\\
11.50$\pm0.25$ & $5$ & $-5.03$ & $-4.58\pm0.26$ & $10.68$\\
\hline
{\bf z = 3.24} & & & & \\
9.75$\pm0.20$ & $50$ & $-3.88$ & $-3.56\pm0.11$ & $9.38$\\
10.15$\pm0.20$ & $49$ & $-3.93$ & $-3.56\pm0.11$ & $10.47$\\
10.55$\pm0.20$ & $19$ & $-4.34$ & $-3.79\pm0.14$ & $10.47$\\
10.95$\pm0.20$ & $6$ & $-4.84$ & $-4.28\pm0.21$ & $10.47$\\
11.35$\pm0.20$ & $1$ & $-5.62$ & $-5.10\pm0.66$ & $10.47$\\
\hline
\end{tabular}}
\label{table:HB_SMF}
\end{table}

\begin{table}
\caption{\oii-selected stellar mass function. Same as in \ref{table:HB_SMF}.}
\resizebox{\columnwidth}{!}{%
\begin{tabular}{c c c c c}
\hline
$\log_{10} M$ & \#  & $\Phi_{\mathrm{obs}} $ & $\Phi_{\mathrm{final}}$ & Volume \\
(M$_\odot$) & & (Mpc$^{-3} ~ \rm{d}\log_{10} \rm{M}$) & (Mpc$^{-3} ~ \rm{d}\log_{10} \rm{M}$) & ($10^5$ Mpc$^{3}$)\\
\hline
{\bf z = 1.47} & & & & \\
9.50$\pm0.20$ & $202$ & $-3.14$ & $-2.97\pm0.09$ & $6.97$\\
9.90$\pm0.20$ & $188$ & $-3.17$ & $-2.97\pm0.09$ & $6.97$\\
10.30$\pm0.20$ & $94$ & $-3.47$ & $-3.22\pm0.10$ & $6.97$\\
10.70$\pm0.20$ & $64$ & $-3.64$ & $-3.35\pm0.11$ & $6.97$\\
11.10$\pm0.20$ & $42$ & $-3.82$ & $-3.46\pm0.11$ & $6.97$\\
11.50$\pm0.20$ & $11$ & $-4.40$ & $-3.97\pm0.16$ & $6.97$\\
11.90$\pm0.20$ & $2$ & $-5.14$ & $-4.32\pm0.35$ & $6.97$\\
\hline
{\bf z = 2.25} & & & & \\
9.60$\pm0.30$ & $47$ & $-3.48$ & $-3.38\pm0.14$ & $2.36$\\
10.20$\pm0.30$ & $43$ & $-3.94$ & $-3.79\pm0.13$ & $6.29$\\
10.80$\pm0.30$ & $22$ & $-4.23$ & $-4.05\pm0.17$ & $6.29$\\
11.40$\pm0.30$ & $7$ & $-4.73$ & $-4.45\pm0.26$ & $6.29$\\
12.00$\pm0.30$ & $1$ & $-5.58$ & $-5.14\pm0.63$ & $6.29$\\
\hline
{\bf z = 3.34} & & & & \\
9.75$\pm0.25$ & $13$ & $-4.44$ & $-4.18\pm0.16$ & $7.13$\\
10.25$\pm0.25$ & $10$ & $-4.73$ & $-4.43\pm0.17$ & $10.84$\\
10.75$\pm0.25$ & $5$ & $-5.14$ & $-4.83\pm0.23$ & $13.81$\\
11.25$\pm0.25$ & $3$ & $-5.36$ & $-5.01\pm0.29$ & $13.81$\\
11.75$\pm0.25$ & $1$ & $-5.84$ & $-5.39\pm0.44$ & $13.81$\\
\hline
\end{tabular}}
\label{table:OII_SMF}
\end{table}

\section{Stellar Mass Comparisons}
\label{sec:comparison}
The COSMOS and UDS fields both have a wealth of multi-wavelength data, which is useful when measuring the physical properties (e.g., stellar masses) of galaxies via SED fitting. Stellar masses for COSMOS includes the $i$-band selected measurements of \citet{Ilbert2010} using {\it Le Phare} and the UltraVISTA/COSMOS $K_s$-band selected measurements of \citet{Muzzin2013} using {\it FAST}. Our \hb- and \oii-selected samples are from both fields but we measure the stellar masses using {\sc MAGPHYS}. This is to ensure that stellar masses are measured using the same SED fitting code in both fields. Not normalizing the stellar mass determinations to the same code can introduce systematic effects arising from model dependencies. 

\begin{figure}
\centering
\includegraphics[width=1.1\columnwidth]{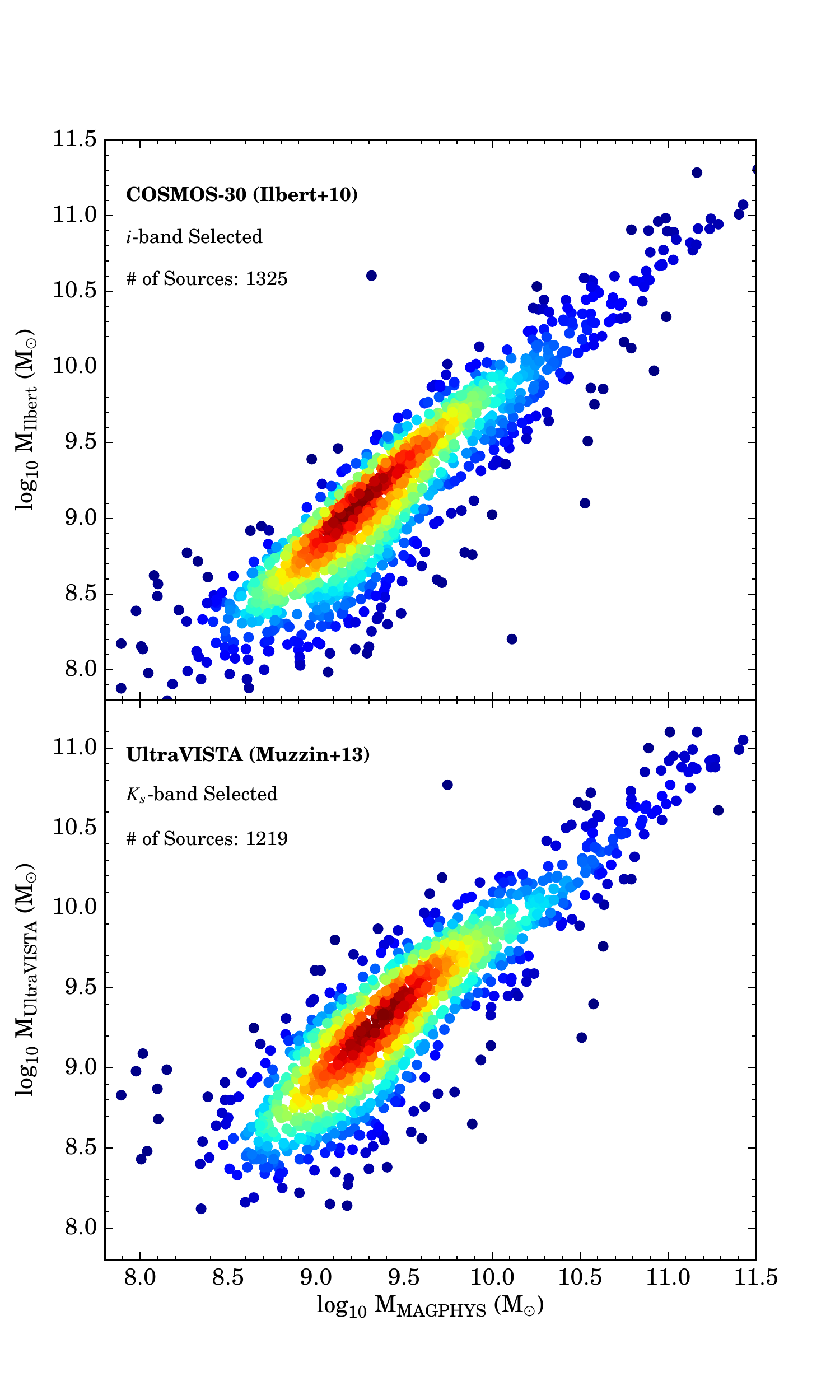}
\caption{Present is the comparison between the stellar masses measured by \citet{Ilbert2010} ({\it top panel}) and \citet{Muzzin2013} ({\it bottom panel}) versus the stellar masses we measure using {\sc MAGPHYS}. We find that, overall, our measurements are consistent with those from the comparison samples. The scatter in the measurements is most probably arising from the different sets of assumptions (e.g., SF history, metallicity range, dust prescription). We eliminate the scatter arising from differing redshifts by only comparing sources that have a $\Delta z < 0.1$, where $\Delta z$ represents the difference between the photometric redshift measured by \citet{Khostovan2015} and the comparison studies.}
\label{fig:SM_compare}
\end{figure}

We compare our stellar mass measurements in Figure \ref{fig:SM_compare} to those of \citet{Ilbert2010} ({\it top panel}) and \citet{Muzzin2013} ({\it bottom panel}). Both studies used a Chabrier IMF but different SED fitting codes and sets of filters, which is the most probable reason for the scatter. To eliminate the scatter arising from redshift differences, we only show comparison measurements for which the difference between the redshift measurement in our catalogs (measured using EaZY, see \citealt{Khostovan2015}) and the comparison measurements is $< 0.1$. Overall, we find that our measurements are consistent with the literature.

\section{Equivalent Width Completeness}
\label{sec:ewr_comp}
We incorporate a second completeness correction which deals with the \ewr~cut causing a loss in high-mass sources (e.g., \S3.1 in \citealt{Sobral2014}; Figure \ref{fig:EW_Mstar}). For our $z > 1$ \hb~and \oii, this incompleteness is not an issue since it will only affect our measurements at very high masses ($>10^{12}$ \msol, except for \oii~$z = 1.47$ where the incompleteness arises by $>10^{11.25}$ \msol). Our $z \sim 0.84$ \hb~sample is affected for masses $> 10^{10}$ \msol. The $z \sim 1.47$ \oii~sample is relatively complete up to masses for which we probe.

We correct for this incompleteness using a similar approach from \citet{Sobral2014}. Since our $z >1$ \hb~are complete for the full range of stellar masses (Figure \ref{fig:hist}), we use these samples as proxies in measuring the incompleteness. We start by binning up the full sample in stellar mass bins which corresponds to a median \ewr. We then decrease the corresponding \ewr~to match the $z \sim 0.84$ median \ewr~stellar mass bins, which results in a number of high-mass sources removed from the full sample because of the $z = 0.84$ \hb~\ewr~cut. The correction factor is then calculated as the number of sources that are recovered relative to the total number of sources in each bin. We run these correction determinations based off the $z \sim 1.47$, $2.23$, and $3.24$ samples and find that for all redshifts probed in the \hb~sample, the \ewr~cut completeness correction does not evolve. To apply the completeness corrections, we extrapolate for the mass range of the $z \sim 0.84$ sample and apply the corrections accordingly. These corrections are mass dependent and range from $\sim 50\%$ to 200\% increase in $\Phi(M)$ between $10^{10}$ \msol~to 200\% and $10^{11.5}$ \msol, respectively.

\section{Correcting $z > 5$ \hb~Equivalent Widths}
\label{sec:correction}
To ensure that the $z > 5$ \hb~literature data points \citep{Labbe2013,Smit2014,Smit2015} are comparable to our measurements, we must take into account the H$\beta$ contribution in the total \ewr~measured. These samples used nebular excess in the {\it Spitzer} IRAC bands to probe the combined H$\beta$4861, [O{\sc iii}]4959, and [O{\sc iii}]5007 lines. Our sample on the other hand consists of either H$\beta$, [O{\sc iii}]4959, or [O{\sc iii}]5007 because the narrow-band filter is narrow enough to separate the lines, but the broad-band and photometric redshift selections used can not. As inferred in \citet{Khostovan2015}, the sample is primarily [O{\sc iii}]5007 for the brightest sources, but towards fainter line fluxes we start picking up more H$\beta$ emitters. \citet{Sobral2015} observed $z = 1.42$ \hb~emitters in the $\sim 10$ deg$^2$ CF-HiZELS survey and had spectroscopic measurements to differentiate between H$\beta$ and \oiii~to find that the sample consisted of primarily \oiii5007 emitters. To compensate for this, we reduce the \ewr~measured by \citet{Labbe2013}, \citet{Smit2014}, and \citet{Smit2015} by 20\% based on the [O{\sc iii}]/H$\beta$ ratios from the $z \sim 2.3$ studies of the MOSDEF survey (e.g., \citealt{Sanders2016}).

\section{Ratios of \ewr~$=$ Fluxes?}
\label{sec:SDSS_VVDS}
The \ioni~ line ratio is typically measured by taking the ratio of the dust-corrected \oiii~and \oii~fluxes. In \S \ref{sec:ion}, we use the ratio of the \ewr~instead of the line fluxes to determine \ioni, as this eliminates the dependency of dust corrections under the assumption that $E(B-V)_\mathrm{nebular} \sim E(B-V)_\mathrm{stellar}$ (e.g., \citealt{Reddy2015,Shivaei2015}). An issue that arises with this approach is that the \ewr~is a ratio between the line and continuum flux, where the continuum flux at 3727\AA~may not be equivalent/similar to the continuum flux at 5007\AA. Therefore, we must assess how well does the \ewr(\oiii)/\ewr(\oii) correlate with $F_{\textrm{\oiii}}/F_\textrm{\oii}$, where the only factor that can cause any systematic deviation is the difference between the continuum fluxes.

To assess this issue, we use the SDSS DR12 \citep{Thomas2013} and VVDS \citep{Lamareille2009} catalogs. Both are spectroscopic surveys and contain the \ewr~and line fluxes for both \oiii~and \oii, allowing us to directly measure the correlation between \ewr(\oiii)/\ewr(\oii) and $F_{\textrm{\oiii}}/F_\textrm{\oii}$. For both catalogs, we select only sources that are confirmed to be star-forming and within the stellar mass range of $9.5 < \log_{10} M < 10.0$ \msol. 

\begin{figure}
\centering
\includegraphics[width=\columnwidth]{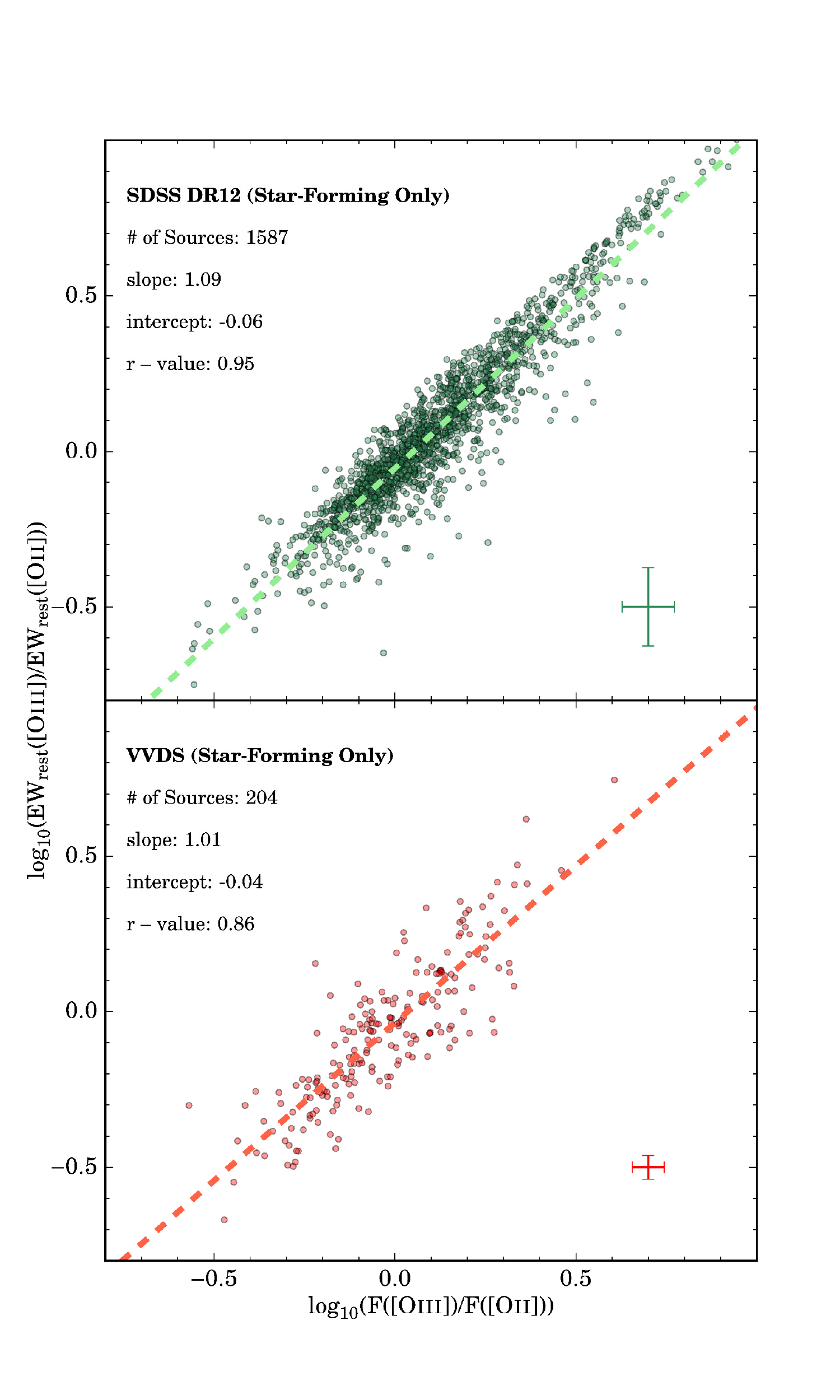}
\caption{The comparison between the \ioni~determined by the ratio of the \ewr~and the line fluxes in log-space. The intercept represents the ratio of the continuum fluxes. We find that the intercepts are $\sim 0$ and the slopes of the correlation are near unity, such that the ratio of the \ewr~directly traces the \ioni~line ratio with negligible systematic offsets introduced by differing continuum fluxes.}
\label{fig:EW_FLUX_comp}
\end{figure}

Figure \ref{fig:EW_FLUX_comp} shows the correlation between \ewr(\oiii)/\ewr(\oii) and $F_{\textrm{\oiii}}/F_\textrm{\oii}$. We measure the correlation in log-scale as:
\begin{equation}
\log_{10} \frac{\mathrm{EW}_\mathrm{rest}(\textrm{\oiii})}{\mathrm{EW}_\mathrm{rest}(\textrm{\oii})} = \log_{10} \frac{F_{\textrm{\oiii}}}{F_\textrm{\oii}} + \log_{10} \frac{f_{C,\textrm{\oii}}}{f_{C,\textrm{\oiii}}}
\end{equation}
where $f_C$ is the continuum flux at the wavelength of the emission line. Therefore, a linear correlation in log-space would have the intercept equivalent to the ratio of the continuum fluxes, which would represent the systematic offset introduced by using the \ewr~ratios to measure \ioni. Furthermore, because we assume $E(B-V)_\mathrm{nebular} \sim E(B-V)_\mathrm{stellar}$ the dust corrections would still cancel out. Changes in this assumption would introduce a systematic factor due to dust correction and not continuum flux differences of $0.4(k_\textrm{\oii} - k_\textrm{\oiii})(E(B-V)_\mathrm{nebular} - E(B-V)_\mathrm{stellar})$ dex.

We find that for the SDSS and VVDS samples, the slope of the correlation is close to unity such that \ewr(\oiii)/\ewr(\oii) $\sim F_{\textrm{\oiii}}/F_\textrm{\oii}$. The $r$-value (correlation coefficient) is $\sim 0.9$ for both samples which implies that the two different ratios are strongly correlated. More importantly, we find that the intercepts measured are $-0.06$ and $-0.04$ dex for SDSS and VVDS, respectively. This suggests that the systematic offset introduced by the ratio of the continuum fluxes is negligible in the determination of \ioni~via the ratio of the \ewr.

\bsp	
\label{lastpage}
\end{document}